\documentclass[%
 aip,
 amsmath,amssymb,
preprint,%
]{revtex4-1}

\usepackage{graphicx}
\usepackage{dcolumn}
\usepackage{bm}

\usepackage[utf8]{inputenc}
\usepackage[T1]{fontenc}
\usepackage{mathptmx}
\usepackage{xcolor}
\usepackage{cleveref}
\usepackage{amsthm}
\usepackage{amssymb} 
\usepackage[caption=false]{subfig}

\newcommand{\AC}{\mathcal{A}}
\newcommand{\BC}{\mathcal{B}}

\newcommand{\FC}{\mathcal{F}}

\newcommand{\WC}{\mathcal{W}}

\newtheorem{theorem}{Theorem}

\newtheorem{definition}[theorem]{Definition}

\begin{document}


\title{Quantum Pontryagin Principle under Continuous Measurements}

\author{J. I. Mulero-Mart\'{\i}nez}
\author{J. Molina-Vilaplana}%
\affiliation{ 
School of Industrial Engineering. Universidad Polit\'ecnica de Cartagena. C/Dr Fleming S/N. 30202. Cartagena. Spain
}%

\date{\today}

\begin{abstract}
In this paper we develop the theory of the quantum Pontryagin principle for continuous measurements and feedback. The analysis is carried out under the assumption of compatible events in the output channel. The plant is a quantum system, which  generally is in a mixed state, coupled to a continuous measurement channel. 
The Pontryagin Maximum Principle is derived in both the Schr\"{o}dinger picture and Heisenberg picture, in particular in statistical moment coordinates. To avoid solving stochastic equations we derive a LQG scheme which is more suitable for control purposes. Finally, we use the quantum harmonic oscillator as a concrete example to illustrate the performance of the controller.
\end{abstract}

\maketitle

\section{Introduction\label{sec:Intro}}

Quantum optimal control (QOC, for short) is a powerful tool for achieving
quantum control objectives in many practical problems of interest for the emerging field of Quantum Technologies ( see \cite{wiseman2010quantum, Glaser2015} and the references therein). It has been successfully used in a vast area of quantum applications from physical chemistry \cite{Shapiro2012}, to multi-dimensional nuclear magnetic resonance experiments \cite{Khaneja2003311}, through time-optimal control problems \cite{PhysRevA.92.063415, PhysRevB.93.035423}.

In constrained optimization, the Pontryagin Maximum Principle (PMP) is an attractive technique based on the variational method of Lagrange multipliers. Few results have appeared in recent years based on the explicit calculation of the extremal solutions and those are applicable only to discrete low-dimensional systems: fast generation of a given-structure wave package 
\cite{Glovinskii}; adiabatic population transfer in $\Lambda $-systems via intermediate states that are subject to decay \cite{PhysRevA.85.033417}; fast adiabatic cooling in harmonic traps and non-interacting collection of harmonic oscillators with a shared frequency \cite{0295-5075-96-6-60015,B816102J,stefanatos11,stefanatos17,stefanatos17b} optimal cooling power of a reciprocating quantum refrigerator; \cite{0295-5075-85-3-30008} time-optimal control for one or two spins (or qubits) \cite{PhysRevA.74.022306,PhysRevA.88.062326,PhysRevLett.104.083001,PhysRevA.63.032308,PhysRevA.88.043422,dalesandro01,boscain02,boscain06,stefanatos10,bonnard09}; minimum-time adiabatic-like paths for the expansion of a quantum piston \cite{Stefanatos20133079}. In those applications the external controls were determined via a
deterministic Pontryagin principle based on the controlled Schr\"{o}dinger
equation:%
\begin{equation*}
\frac{d}{dt}\, \left\vert \psi\right\rangle =-\frac{i}{\hbar}\, H\left(u\right) \left\vert
\psi \right\rangle
\end{equation*}%
where $H\left( u\right) $ is the Hamiltonian driven by a control action $u$ and $\left\vert \psi \right\rangle $ is a state vector in the Hilbert space of the system to be controlled. In most of these aforementioned approaches, the feedback is absent \cite{roloff09}. In other words, the system is not subjected to any measurement and a subsequent {\it quantum filtering}, which reduces the Pontryagin principle to its deterministic version (see \cite{sugny08, wang11,egger14} for examples and applications of quantum optimal control with measurements). On the other hand, a few results promote ad-hoc model-dependent solutions\cite{Glovinskii}.

Motivated by these previous studies, here we address the task of stating both a firm theoretical ground and a formalization of the quantum PMP for systems with quantum feedback provided by continuous weak measurements \cite{Jacobs06,Clerk10}. This may help the subsequent verification procedures to check if the resulting control is optimal \cite{Jacobs08}. In addition, a quantum Pontryagin principle would allow to tackle with state constraints. 

In the Schr\"{o}dinger picture the conditional dynamics of the system evolving under these weak measurements is described by quantum filtering theory. Filtering concerns the processing of the information yielded by the measurement process. This information is generally incomplete because the state is not fully accessible by the measurement setting, and is inherently corrupted by noise. In this context, optimal control problems are solved by using a cost function expressed in terms of the state given by the filter, which is often called an information state.  The quantum Belavkin filter, or stochastic master equation, computes this information state \cite{belavkin99}.

This work is organized as follows. In Section \ref{sec:QOCP} the continuous measurement of a quantum system  and the subsequent filtering of the outcomes of this measurement
is analyzed in terms of a quantum probability space. There, measurement operators are considered for the abelian subalgebra generated by commuting projectors. As a result, filtrations in the quantum probability space as well as adapted processes are defined in terms of von Neumann subalgebras. This allows us to define feasible and admissible control actions in terms of operators. Section \ref{sec:QPPSP} is devoted to the derivation of the quantum PMP in a global form from the Hamilton-Jacobi-Bellman equation in the Schr\"{o}dinger picture. The result is a system of coupled forward-backward stochastic differential equations that replaces the HJB partial differential equation. 

In addition, as a very natural way of defining state-space realizations is in the Heisenberg picture, particularly in coordinates of expectation and variance, we derive in \ref{sec:QPPHP} a Hamilton-Jacobi-Bellman equation in these coordinates as a preceding step to build a quantum PMP. One drawback of stochastic PMP is that the control problem cannot be solved in closed form, and one needs to resort to numerical optimization. To overcome this difficulty, for quantum linear models \cite{petersen16} we derive a LQG from the quantum PMP based on the statistical moment coordinates. This adds a novel result to the significant literature on measurement based optimal control of quantum linear systems (see \cite{dong09,nori17} and references therein). In Section \ref{sec:Example} we illustrate the application of the quantum PMP to a dissipative quantum harmonic oscillator. Finally, the concluding remarks are given in Section \ref{sec:Conclusions}.

\section{Quantum Optimal Control Problem}
\label{sec:QOCP}
As we formulate the quantum Pontryagin's Principle under continuous measurements, first we review how the  quantum measurement and 
filtering problem can be rephrased in terms  of the  optimal estimation of the output of a noisy quantum channel. 

With every quantum system there is a separable complex Hilbert space $\mathcal{H}$ on which a von Neumann algebra of linear operators $\mathcal{A}$ is defined (a Banach algebra of bounded operators on $\mathcal{H}$). In general, one also needs the predual space $\mathcal{A}_{\star }$ and particularly those positive elements of $\mathcal{A}_{\star }$ which are unitary normed, called density operators $\rho$ or normal states. A quantum probability space can be defined as the pair $\left( \mathcal{A},\rho \right) $.

In this paper, quantum feedback is generated through a continuous quantum measurement on a quantum system. Following \cite{belavkin99}, this continuous measurement is implemented by an indirect measurement of the operators corresponding to a semi-classical field coupled to the system. After measurement, the change of the density operator of the system  $\rho$ is described through an optimal estimator based on the results of measurements in this field. The state of the field lies in the  Hilbert space $\FC$ and initially is in its vacuum state $\phi$.  We observe compatible events on this measurement channel through projectors $\{P_\omega\}_{\omega\in\Omega}$ that generate an abelian subalgebra $\BC\subset\BC(\FC)$ where $\Omega$ is  the space of measurement results (eigenvalues) for these commuting operators.
In this sense, all the field operators $W\in\BC$ which are linear combinations
of the commuting projectors are in one-to-one correspondence
with classical random variables as functions from the data space $\Omega$ into $\mathbb{R}$.  
The interaction between the quantum system and the field is described in terms of 
a unitary operator $U$ such that, 
\begin{displaymath}
\rho\otimes \phi\to U(t)(\rho\otimes\phi)U^{\dagger}(t)\, .
\end{displaymath}

With this, the conditional evolution after a measurement on the quantum system is given by 
\begin{equation}\label{eq poststate}
\rho(t)\to\rho_\omega(t)\:=\frac{\textrm{Tr}_\FC[U(t)(\rho\otimes\phi)U^{\dagger}(t)(I\otimes
P_\omega)]}{\textrm{Tr}[U(t)(\rho\otimes\phi)U^{\dagger}(t)(I\otimes P_\omega)]}
\end{equation}
 which gives the \emph{posterior state} $\rho_{\omega}(t)$ that implements the Bayes law of conditioning for the measurement result $\omega\in\Omega$,
normalized with respect to the output probabilities $\mathbb{P}(\omega)=\textrm{Tr}[U(t)(\rho\otimes\phi)U^{\dagger}(t)(I\otimes P_\omega)]$
and $\textrm{Tr}_\FC$ denotes the partial trace over $\FC$.

The posterior state must be considered as a classical random variable
$\rho_\bullet:\Omega\to\AC_{\star}$ taking values $\rho_\omega$
in the space $\AC_{\star}$ of states on $\AC$.
It gives the conditional expectation
\begin{displaymath}
\mathbb{E}[X'|Y]=\langle\rho_\bullet,X\rangle
\end{displaymath}
which amounts to the least squares estimator of the system operator $X'_t=U^{\dagger}(t)(X\otimes I)U(t)$
after interaction, with respect to the output operators $Y_t:=U^{\dagger}(t)(I\otimes W)U(t)$.

Within this framework, the dynamical coupling between the system and the field acts 
as a quantum noise bath on the system. The bath is then modelled by a Fock space
$\FC$ and $\WC:=\BC(\FC)$ is the algebra of bounded operators on $\FC$. 
Thus, the continuous measurement of our system is understood as measurements of 
a Wiener process in the field described by field quadrature operators $W_t=A_t+A_t^{\dagger}$
where $A_t\in \WC$ is the \emph{annihilation} operator
on $\FC$. Quantum stochastic calculus can be defined using the annihilation
process and its adjoint, the creation process
$A^{\dagger}_t$ as the fundamental diffusive adapted
processes whose increments $dA_t$, $dA^{\dagger}_t$
are considered as operators acting in the Fock space by
using the multiplication table \cite{DBLP:journals/siamco/BoutenHJ07,parthasarathy2012introduction}
\begin{equation}\label{eq mult}\begin{array}{l}
(dt)^2=0,\quad dtdA_t=0=dtdA^{\dagger}_t, \\dA_t^{\dagger}dA_t=0,\quad dA_tdA_t^{\dagger}=dt.
\end{array}\end{equation}

Denoting $L$ as the operator that models the coupling of the system to the measurement channel, a time continuous measurement of $W_t$ in the output
channel represents an indirect measurement
of $L_t+L_t^{\dagger}\in\AC_t$ as can be seen from
the quantum It\^o formula applied
to the directly observable output operators $Y_t=U(t)^{\dagger}(I\otimes
W_t)U(t)$:
\begin{equation}\label{eq flowY}
dY_t=(L_t+L^{\dagger}_t)\otimes dt+I\otimes dW_t\, .
\end{equation}

In order  to estimate the value of dynamical variables for which only an incomplete knowledge is provided through the indirect kind of oservations described above, some filtering equations are needed. Belavkin was the first to provide the quantum
filtering equation which describes the optimal estimate of a density matrix conditioned by the classical output of a noisy quantum channel. 
The conditional expectation $\mathbb{E}[X_t|Y_t]$
gives the least squares estimator $\hat{X}_t$ of
a conditional operator $X_t\in \AC_t$ on the output operators
$Y_t$. This is equivalent to a classical random variable on the space of
measurement trajectories $\Omega_T:=\{\omega_\tau|0\le t \le T$ s.t.
$\omega_t$ is an eigenvalue of $Y_t\}$ with $T\in \left( 0,\infty \right) $ fixed.
This conditional expectation is most conveniently written in the Schr\"odinger
picture $\mathbb{E}[X_t|Y_t]=\langle \rho_\bullet(t),X\rangle$ for the solution
$\rho_\bullet(t)$ to the stochastic nonlinear Schr\"{o}dinger-like differential equation
\begin{equation}
d\rho _{\bullet }\left( t\right) =w\left( t,\rho _{\bullet
}\left( t\right) \right) dt+\sigma \left( \rho _{\bullet }\left( t\right)
\right) dW_t  \label{EQ1}
\end{equation}%
called the Belavkin quantum filter in which Ito calculus  $dW_t\, dW_t= dt$ holds. In Eq.$(\ref{EQ1})$ $w:\left[ 0,T\right] \times \mathcal{S}\rightarrow \mathcal{S}$, with $\mathcal{S}\subset \mathcal{A}_{\star }$ the convex set of density operators, refers to the unconditional evolution of states given by the dual of a Lindblad generator $\mathcal{L}$:  
\begin{equation*}
w\left( t,\rho_\bullet(t)\right) =\mathcal{L}^{\dagger }\left( \rho_\bullet(t) \right) =-\frac{i}{%
\hbar }\left[ H,\rho_\bullet(t) \right] +\frac{1}{2}\left( L\left[ \rho_\bullet(t) ,L^{\dagger }%
\right] +\left[ L,\rho_\bullet(t) \right] L^{\dagger }\right)\, ,
\end{equation*}%
and the fluctuation operator $\sigma :\mathcal{S}\rightarrow \mathcal{S}$ is
given by, 
\begin{displaymath}
\sigma(\rho_\bullet(t))=\rho_\bullet(t) L^{\dagger}+L\rho_\bullet(t)-
\langle\rho_\bullet(t),L^{\dagger}+L\rangle
\rho_\bullet(t)\, .
\end{displaymath}

Equation (\ref{EQ1}) defines a filtered quantum probability space $\left\{ \left( \mathcal{A},\rho ,%
\mathcal{A}_{t]}\right) ,t\geq 0\right\} $ satisfying the usual condition (
i.e. $\left( \mathcal{A},\rho \right) $ is complete, $\mathbf{I}\in \mathcal{%
A}_{0]}$ and $\left\{ \mathcal{A}_{t]},t\geq 0\right\} $ is right
continuous).

By the information field $\left\{ \mathcal{A}_{t]},t\geq 0\right\} $, the
controller is well-informed of what has happened in the past but, because of
the uncertainty of the system, it is not able to predict the future. As a
consequence there exists a non-anticipative restriction on the controller:
for any instant the controller cannot decide its control action before the
instant occurs. This restriction is expressed in mathematical terms as "$%
u\left( \cdot \right) $ is $\left\{ \mathcal{A}_{t]}\right\} _{t>0}$
adapted", and the control is taken from the set%
\begin{equation*}
\mathcal{U}_{\left[ 0,T\right] }=\left\{ u:\Omega _{T}\times \left[ 0,T%
\right] \rightarrow U:u\left( \cdot \right) \text{ is }\left\{ \mathcal{A}%
_{t]}\right\} _{t>0}\text{ adapted}\right\}
\end{equation*}%
Any $u_{\bullet }\left( \cdot \right) \in \mathcal{U}_{\left[ 0,T\right] }$
is called a feasible control. In particular fixing a control the drift and the dispersion
depend on $\Omega_{T}$, i.e. $w,\sigma:\Omega_{T}\times\mathbb{R}%
^{+}\rightarrow\mathcal{A}_{\star}$. Given a $\mathcal{A}_{0]}$-measurable
density matrix $\rho$ for every $u_{\bullet}\left(  \cdot\right)  $ the master
equation \ref{EQ1} admits a strong solution $\rho\left(  t;u\right)  $ ( seen as a
continuous $\left\{  \mathcal{A}_{t]},t\geq0\right\}  -$adapted process) if 

(i) $\rho\left(  0\right)$ is almost sure in probability.

(ii) $\int_{0}^{t}\left(  \left\Vert w\left(  \tau,\rho_{\omega}\left(
\tau\right)  ,u_{\omega}\left(  \tau\right)  \right)  \right\Vert +\left\Vert
\sigma\left(  \tau,\rho_{\omega}\left(  \tau\right)  \right)  \right\Vert
^{2}\right)  d\tau<\infty$ for $t\geq0$, almost sure in probability for
$\omega\in\Omega_{T}$.

(iii) $\rho\left(  t\right)  =\rho+\int_{0}^{t}w\left(  \tau,\rho,u\right)
d\tau+\int_{0}^{t}\sigma\left(  \tau,\rho\right)  dW\left(  \tau\right)  $,
$t\geq0$, almost sure in probability\newline The adapted solution exists and
it is unique if the drift and the dispersion are continuous measurable
functions and they satisfy a Lipschitz condition with respect to $\rho$.

The adapted solution exists and
it is unique if the drift and the dispersion are continuous measurable
functions and they satisfy a Lipschitz condition with respect to $\rho$.

Let $\rho _{\bullet }\left( \cdot \right) $ be
the solution to the filtering equation with initial condition $\rho
_{\bullet }\left( 0\right) =\rho $. The cost $\emph{J}$ for a feasible
control action $u_{\bullet }\left( \cdot \right) $ is a random variable on $%
\Omega _{T}$, i.e. $\emph{J}:\Omega _{T}\times \mathcal{U}_{\left[ 0,T\right]
}\times \mathcal{S}\times \left[ 0,T\right] \rightarrow \mathbb{R}$, defined
by%
\begin{equation}
\emph{J}_{\bullet }\left( u_{\bullet }\left( \cdot \right) ;\rho _{\bullet
}\left( \cdot \right) ,t\right) =\int_{0}^{T}\emph{C}\left( \tau ,u_{\bullet
}\left( \tau \right) ,\rho _{\bullet }\left( \tau \right) \right) d\tau +%
\emph{M}\left( \rho _{\bullet }\left( T\right) \right)  \label{EQ17}
\end{equation}%
where the cost density $\emph{C}$ and the terminal cost $\emph{M}$ are
linear; for each $\omega \in \Omega _{T}$ and each instant $\tau \in \left(
0,T\right] $:%
\begin{eqnarray*}
\emph{C}\left( u_{\omega }\left( \tau \right) ,\rho _{\omega }\left( \tau
\right) \right) &=&\left\langle \rho _{\omega }\left( \tau \right) ,C\left(
\tau ,u_{\omega }\left( \tau \right) \right) \right\rangle \\
\emph{M}\left( \rho _{\omega }\left( T\right) \right) &=&\left\langle \rho
_{\omega },M\right\rangle
\end{eqnarray*}%

with $C$ and $M$ positive self-adjoint operators. Note that this cost
may be different according to the requirements of the QOC problem, such as
minimizing the control time, the control energy, the error between the final
state and target state, or a combination of these. The goal is to minimize
the criterion by selecting a nonanticipative decision among the ones
satisfying all the quantum state constraints:

\begin{definition}
Let $\left\{ \left( \mathcal{A},\rho ,\mathcal{A}_{t]}\right) ,t\geq
0\right\} $ be a filtered quantum probability space with the usual condition
and let $W $ be a given standard $\left\{ \mathcal{A}%
_{t]},t\geq 0\right\} -$Wiener process. A control $u_{\bullet }\left( \cdot
\right) $ is called q-admissible, and $\left( \rho _{\bullet }\left( \cdot
\right) ,u_{\bullet }\left( \cdot \right) \right) $ is a q-admissible pair if%
\newline
(i) $u_{\bullet }\left( \cdot \right) \in \mathcal{U}_{\left[ 0,T\right] }$.%
\newline
(ii) $\rho _{\bullet }\left( \cdot \right) $ is the unique solution to the
master equation $\left( \ref{EQ1}\right) $.\newline
(iii) $\rho _{\bullet }\left( \cdot \right) $ satisfies a prescribed state
constraint.\newline
(iv) For each $\omega \in \Omega _{T}$, $\emph{C}\left( \cdot ,u_{\omega
}\left( \tau \right) ,\rho _{\omega }\left( \tau \right) \right) \in L_{%
\mathcal{A}}^{1}\left( 0,T;\mathbb{R}\right) $, $\emph{S}\left( \rho
_{\omega }\left( T\right) \right) \in L_{\mathcal{A}_{T}}^{1}\left( \Omega
_{T};\mathbb{R}\right)$.
\end{definition}

Here the spaces $L_{\mathcal{A}}^{1}\left( 0,T;\mathbb{R}\right)$ and $L_{\mathcal{A}_{T}}^{1}\left( \Omega
_{T};\mathbb{R}\right)$ are defined on the filtered probability space $\left\{ \left( \mathcal{A},\rho ,%
\mathcal{A}_{t]}\right) ,t\geq 0\right\} $:
$L_{\mathcal{A}}^{1}\left( 0,T;\mathbb{R}\right)$ is the set of all $\left\{  \mathcal{A}_{\left.
t\right]  }\right\}  _{t\geq0}$-adapted $\mathbb{R}$-valued processes $X\left(
\bullet\right)  $ such that $\mathbb{E}\left[  \int_{0}^{T}\left\vert X\left(
t\right)  \right\vert dt\right]  <\infty$ and $L_{\mathcal{A}_{T}}^{1}\left( \Omega
_{T};\mathbb{R}\right)$ is the set of $\mathbb{R}$-valued $\mathcal{A}_{T}$-measurable random variables
$X$ such that $\mathbb{E}\left[  \left\vert X\right\vert \right]  <\infty$.

The filtration $\left\{ \mathcal{A}_{t]},t\geq 0\right\} $ as well as the
Wiener process $W$ are fixed and independent of the
control. The set of all q-admissible controls will be denoted by $\mathcal{U}%
_{ad}\left[ 0,T\right] $. The quantum optimal control (QOC) problem can be
stated as follows:

\textbf{Problem QOC}: Minimize $\mathbb{E}\left[ \emph{J}_{\bullet }\left(
u_{\bullet }\left( \cdot \right) ;\rho _{\bullet }\left( \cdot \right)
,t\right) \right] $ over $\mathcal{U}_{ad}\left[ 0,T\right] $, where $%
\mathbb{E}\left[ \cdot \right] $ denotes the expectation value on $\Omega_{T}$, i.e. the expectation over all the possible continuous
trajectories $\omega_t  $ with $0\leq t\leq T$.

The goal is to find $u_{\bullet }^{\ast }\left( \cdot \right) \in \mathcal{U}%
_{ad}\left[ 0,T\right] $ such that%
\begin{equation}
\mathbb{E}\left[ \emph{J}_{\bullet }\left( u_{\bullet }^{\ast }\left( \cdot
\right) \right) \right] =\min_{u_{\bullet }\left( \cdot \right) \in \mathcal{%
U}_{ad}\left[ 0,T\right] }\mathbb{E}\left[ \emph{J}_{\bullet }\left(
u_{\bullet }\left( \cdot \right) \right) \right]  \label{EQ2}
\end{equation}%
The control task can be formulated as a problem of searching for a set of
admissible controls satisfying the system dynamic equations while
simultaneously minimizing a cost functional. Any $u_{\bullet }^{\ast }\left(
\cdot \right) \in \mathcal{U}_{ad}\left[ 0,T\right] $ satisfying $\left( \ref%
{EQ2}\right) $ is called a q-optimal control and if it is unique the problem
QOC is said to be q-solvable. The corresponding state process $\rho
_{\bullet }^{\ast }\left( \cdot \right) $ and the state-control pair $\left(
u_{\bullet }^{\ast }\left( \cdot \right) ,\rho _{\bullet }^{\ast }\left(
\cdot \right) \right) $ will be called a q-optimal state process and a
q-optimal pair, respectively.

\section{Quantum Pontryagin Principle in the Schr\"{o}dinger Picture}
\label{sec:QPPSP}

The objective of this section is to derive the quantum Pontryagin's Maximum
Principle in a global form from the Hamilton-Jacobi-Bellman equation (HJB)
for quantum optimal control. The result is a system of coupled
forward-backward stochastic differential equations that replaces the HJB
partial differential equation.

An admissible control $u_{\bullet }\left( \cdot \right) $ is $\left\{ 
\mathcal{A}_{t]},t\geq 0\right\} $-adapted implies that the quantum state $%
\rho $ is actually not uncertain for the controller at time $t$, and this
means that $\rho $ is almost surely deterministic under an appropriate
probability measure\footnote{The probability measure can be defined as $\mathbb{P}\left(
\left.  \cdot\right\vert \mathcal{A}_{s]}\right)  \left(  \omega\right)  $ for
a fixed $\omega\in\Omega$. For any $t\in\left[  0,s\right]  $,%
\begin{gather}
\mathbb{P}\left(  \left.  \left\{  \omega^{\prime}\in\Omega:u\left(
t,\omega^{\prime}\right)  =u\left(  t,\omega\right)  \right\}  \right\vert
\mathcal{A}_{s]}\right)  \left(  \omega\right)  =\\
=\mathbb{E}\left[  \left.  I_{\left\{  \omega^{\prime}\in\Omega:u\left(
t,\omega^{\prime}\right)  =u\left(  t,\omega\right)  \right\}  }\right\vert
\mathcal{A}_{s]}\right]  \left(  \omega\right)  
\end{gather}
where $I_{\left\{  \omega^{\prime}\in\Omega:u\left(  t,\omega^{\prime}\right)
=u\left(  t,\omega\right)  \right\}  }$ is the indicator function. It is
obvious that%
\[
\mathbb{E}\left[  \left.  I_{\left\{  \omega^{\prime}\in\Omega:u\left(
t,\omega^{\prime}\right)  =u\left(  t,\omega\right)  \right\}  }\right\vert
\mathcal{A}_{s]}\right]  \left(  \omega\right)  =I_{\left\{  \omega^{\prime
}\in\Omega:u\left(  t,\omega^{\prime}\right)  =u\left(  t,\omega\right)
\right\}  }\left(  \omega\right)  =1
\]

Therefore, the event $\left\{  \omega^{\prime}\in\Omega:u\left(
t,\omega^{\prime}\right)  =u\left(  t,\omega\right)  \right\}  $ happens
almost surely. This means that for a fixed $\omega$, the action control $u\left(  t \right)  $ is almost surely a deterministic constant $u\left(t,\omega\right)  $ for any $t\leq s$.}
\newline Given a q-optimal control $u_{\bullet }^{\ast }\left(
\cdot \right) $, let us denote $S\left( t,\rho \right) $ as a minimum
posterior cost-to-go ( sometimes called the value function): 
\begin{equation*}
S\left( t,\rho \right) =\min_{u_{\bullet }\left( \cdot \right) \in \mathcal{U%
}_{ad}\left[ t,T\right] }\mathbb{E}\left[ J_{\bullet }\left( t,\rho
;u_{\bullet }\left( \cdot \right) \right) \right] \text{ }\forall t\in \left[
0,T\right) \times \mathcal{S}
\end{equation*}%
which solves the quantum HJB equation derived from the quantum filtering
equation for all $\left( t,\rho \right) \in \left[ 0,T\right) \times 
\mathcal{S}$:%
\begin{equation}
-\frac{\partial S\left( t,\rho \right) }{\partial t}=\min_{u\in U}\mathcal{H}%
\left( t,u,\rho ,\frac{\partial S\left( t,\rho \right) }{\partial \rho }%
,\left( \frac{\partial }{\partial \rho }\otimes \frac{\partial }{\partial
\rho }\right) S\left( t,\rho \right) \right)  \label{EQ5}
\end{equation}%
where $\mathcal{H}$ is the generalized Hamiltonian%
\begin{gather}
\mathcal{H}\left( t,u,\rho ,p,P\right) =\frac{1}{2}\left\langle \sigma
\left( \rho \right) \otimes \sigma \left( \rho \right) ,P\right\rangle 
\notag \\
-\left\langle w\left( t,u,\rho \right) ,p\right\rangle +\emph{C}\left(
t,u,\rho \right)
\end{gather}%
The operators $p\left(  t,\rho\right)  $ and $P\left(  t,\rho\right)  $
are defined as follows%
\begin{align*}
p\left(  t,\rho\right)    & =\frac{\partial}{\partial\rho}\mathcal{S}\left(
t,\rho\right)  \\
P\left(  t,\rho\right)    & =\left(  \frac{\partial}{\partial\rho}\otimes
\frac{\partial}{\partial\rho}\right)  \left(  \mathcal{S}\left(
t,\rho\right)  \right)
\end{align*}
The operational differentiation of $\mathcal{S}:\mathbb{R}^{+}\times
\mathcal{A}_{\star}\rightarrow\mathbb{R}$ with respect to the density matrix
$\rho$ is computed in a natural way:%
\[
\frac{\partial\mathcal{S}}{\partial\rho}=\left(
\begin{array}
[c]{ccc}%
\frac{\partial\mathcal{S}}{\partial\rho_{11}} & \cdots & \frac{\partial
\mathcal{S}}{\partial\rho_{1n}}\\
\vdots & \ddots & \vdots\\
\frac{\partial\mathcal{S}}{\partial\rho_{n1}} & \cdots & \frac{\partial
\mathcal{S}}{\partial\rho_{nn}}%
\end{array}
\right)
\]
and
\[
\left(  \frac{\partial}{\partial\rho}\otimes\frac{\partial}{\partial\rho
}\right)  \left(  \mathcal{S}\left(  t,\rho\right)  \right)  =\left(
\begin{array}
[c]{ccc}%
\frac{\partial^{2}\mathcal{S}}{\partial\rho\partial\rho_{11}} & \cdots &
\frac{\partial^{2}\mathcal{S}}{\partial\rho\partial\rho_{1n}}\\
\vdots & \ddots & \vdots\\
\frac{\partial^{2}\mathcal{S}}{\partial\rho\partial\rho_{n1}} & \cdots &
\frac{\partial^{2}\mathcal{S}}{\partial\rho\partial\rho_{nn}}%
\end{array}
\right)
\]
In the master equation the fluctuation operator $\sigma $ does not depend of
the control action and is not degenerated. At any time instant the
controller is knowledgeable about some information (as specified by the
information field $\left\{ \mathcal{A}_{t]},t\geq 0\right\} $ of what has
occurred up to that moment, but not able to predict what is going to happen
afterwards due to the uncertainty of the system. It is remarkable that $\left\{  \mathcal{A}_{t]},t\geq0\right\}  $ is the
natural filtration generated by $W$ so that the unique source of uncertainty
proceeds from the noise of the measurement and all the past information around
the noise is available to the controller. Let $u^{\ast }\left( \cdot
\right) $ be the optimal control action in the convex control domain $%
\mathcal{U}$.

From the definition of $p$ and $\ P$ it is obvious that 
\begin{equation*}
\mathcal{H}\left( t,u^{\ast },\rho ,p,q\right) =\inf_{u\left( \cdot \right)
\in \mathcal{U}}\mathcal{H}\left( t,u,\rho ,p,P\right)
\end{equation*}%
The quantum filtering equation together with the optimal control can be
written in terms of $\mathcal{H}$:%
\begin{equation}
d\rho \left( t\right) =-\frac{\partial \mathcal{H}}{\partial p}\left(
t,u^{\ast },p,P\right) dt+\sigma \left( \rho \right) dW,  \label{EQ9}
\end{equation}%
where from here in advance for short $dW_t \equiv dW$ is used. 
Given that the adjoint operator $p$ depends both on time and the density
operator $\rho $, we can compute its differential resorting to It\^{o}'s
lemma:%
\begin{equation*}
dp=\frac{\partial p}{\partial t}dt+\left\langle \frac{\partial }{\partial
\rho },d\rho \right\rangle \left( p\right) +\frac{1}{2}\left\langle \frac{%
\partial }{\partial \rho }\otimes \frac{\partial }{\partial \rho },d\rho
\otimes d\rho \right\rangle \left( p\right)
\end{equation*}%
where the inner products are differential operators applied to the adjoint
variable $p$; specifically $\left\langle \cdot,\cdot\right\rangle $ stands for the
Frobenius inner product, for example%
\[
\left\langle \frac{\partial}{\partial\rho},d\rho\right\rangle =\sum
_{i,j=1}^{n}\frac{\partial}{\partial\rho_{i,j}}d\rho_{i,j}%
\]
and%
\[
\left\langle \frac{\partial}{\partial\rho}\otimes\frac{\partial}{\partial\rho
},d\rho\otimes d\rho\right\rangle =\sum_{i,j,k,l=1}^{n}\frac{\partial
}{\partial\rho_{i,j}}\frac{\partial}{\partial\rho_{k,l}}d\rho_{i,j}d\rho_{k,l}%
\] 
The tensor product $d\rho \otimes d\rho $ can
be rewritten from $\left( \ref{EQ9}\right) $ by noting that $\frac{\partial 
\mathcal{H}}{\partial p}\left( t,u^{\ast },p,P\right) =-w\left( t,u^{\ast
},\rho \right) $:%
\begin{gather*}
d\rho \otimes d\rho =\left( w\otimes w\right) dt^{2}+\left( w\otimes \sigma
\left( \rho \right) dW\right) dt \\
+\left( \sigma \left( \rho \right) dW\otimes w\right) dt+\left( \sigma
\left( \rho \right) \otimes \sigma \left( \rho \right) \right) 
dW^2
\end{gather*}%
For $dt\rightarrow 0$ we use the It\^{o}'s rules: $dt^{2}\rightarrow 0$, $%
dWdt\rightarrow 0$, $ dW^2 \rightarrow dt$. Thus $d\rho
\otimes d\rho =\left( \sigma \left( \rho \right) \otimes \sigma \left( \rho
\right) \right) dt$ and%
\begin{equation}
dp=\frac{\partial p}{\partial t}dt+\left\langle \frac{\partial }{\partial
\rho },d\rho \right\rangle \left( p\right) +\frac{1}{2}\left\langle \frac{%
\partial }{\partial \rho }\otimes \frac{\partial }{\partial \rho },\sigma
\left( \rho \right) \otimes \sigma \left( \rho \right) \right\rangle \left(
p\right) dt  \label{EQ10}
\end{equation}%
Inserting $d\rho $ into $\left( \ref{EQ10}\right) $ yields%
\begin{gather*}
dp=\frac{\partial p}{\partial t}dt+\left\langle \frac{\partial }{\partial
\rho }\left( \cdot \right) ,\frac{\partial \mathcal{H}}{\partial p}\left(
t,u,p,P\right) dt\right\rangle \left( p\right) dt \\
+\left\langle \frac{\partial }{\partial \rho }\left( \cdot \right) ,\sigma
\left( \rho \right) dW\right\rangle \left( p\right) \\
+\frac{1}{2}\left\langle \frac{\partial }{\partial \rho }\left( \cdot
\right) \otimes \frac{\partial }{\partial \rho }\left( \cdot \right) ,\sigma
\left( \rho \right) \otimes \sigma \left( \rho \right) \right\rangle \left(
p\right) dt
\end{gather*}%
Given that $p\left( t,\rho \right) =\frac{\partial S\left( t,\rho \right) }{%
\partial \rho }$ it follows that $\frac{\partial p}{\partial t}=\frac{%
\partial }{\partial \rho }\frac{\partial S\left( t,\rho \right) }{\partial t}
$, and from $\left( \ref{EQ5}\right) $: 
\begin{eqnarray*}
-\frac{\partial p}{\partial t} &=&\frac{\partial }{\partial \rho }\mathcal{H}%
\left( t,u^{\ast },\rho ,p,P\right) =\frac{\partial \mathcal{H}}{\partial
\rho }+\left\langle \frac{\partial \mathcal{H}}{\partial p},\frac{\partial }{%
\partial \rho }\left( \cdot \right) \right\rangle \left( p\right) + \\
&&+\left\langle \frac{\partial \mathcal{H}}{\partial q},\frac{\partial }{%
\partial \rho }\left( \cdot \right) \otimes \frac{\partial }{\partial \rho }%
\left( \cdot \right) \right\rangle \left( p\right)
\end{eqnarray*}%
As a result,%
\begin{equation}
dp=-\frac{\partial \mathcal{H}}{\partial \rho }dt+\left\langle \frac{%
\partial }{\partial \rho }\left( \cdot \right) ,\sigma \left( \rho \right)
dW\right\rangle \left( p\right)
\label{EQ1000}
\end{equation}%
The last equation can be rewritten in a compact form by defining the first order adjoint operator 
\begin{equation*}
q\left(
t,\rho\right)=%
\left\langle \frac{\partial }{\partial \rho }\left( \cdot \right) ,\sigma
\left( \rho \right)\right\rangle \left( \frac{\partial S\left(
t,\rho \right) }{\partial \rho }\right)
\end{equation*}%
The evolution of the adjoint operator $p\left(
t,\rho\right)$ is given by%
\begin{equation}
dp=-\frac{\partial \mathcal{H}}{\partial \rho }dt+q\left(
t,\rho\right) %
 dW  \label{EQ3}
\end{equation}%
The first order adjoint system $\left(  p\left(  t,\rho\right)  ,q\left(
t,\rho\right)  \right)  $ is a pair of $\left\{  \mathcal{A}_{t]}%
,t\geq0\right\}  -$adapted processes which give a solution of the backward
stochastic differential equation $\left( \ref{EQ3}\right)  $. Furthermore every pair
satisfying equation $\left( \ref{EQ3}\right)  $ is an adapted solution. The adjoint
operators  $p,q\in L_{\mathcal{A}}^{2}\left(  0,T;\mathcal{A}_{\star}\right)
$ now live in the same space and have the same dimension.
We are in order to write the systems of forward-backward quantum differential
equations for the Quantum Pontryagin principle:%
\begin{eqnarray*}
d\rho ^{\ast } &=&-\frac{\partial \mathcal{H}^{\ast }}{\partial p}dt+\sigma
^{\ast }\left( \rho ^{\ast }\right) dW \\
dp^{\ast } &=&-\frac{\partial \mathcal{H}^{\ast }}{\partial \rho }dt+q
^{\ast }\left( \rho ^{\ast }\right) dW \\
\rho ^{\ast }\left( 0\right) &=&\rho _{0} \\
p^{\ast }\left( T\right) &=&\frac{\partial S\left( T,\rho ^{\ast
}\left( T\right) \right) }{\partial \rho }
\end{eqnarray*}%
where the superscript '$\ast $' indicates that the terms are optimal.

Here an explicit equation with boundary conditions for the first order adjoint
operator $q$ is not necessary since $q$ is connected to $p$ in a differential way.

\section{Quantum Pontryagin Principle in the Heisenberg Picture}
\label{sec:QPPHP}

\subsection{Quantum Linear Model}

We make the following standard assumptions, \cite{petersen16,gardiner2004quantum}:

\textbf{(A1)} The environment ( or thermal bath) is described by a
quantized electromagnetic field, i.e. a collection of quantum harmonic
fields each of them corresponding to a mode of the field at a given angular
frequency. The interaction between the system and the environment admits a
field interpretation as a transmission line.

\textbf{(A2)} We assume the rotating wave approximation i.e. the neglection
of highly-oscillating terms in the energy flowing between the system and the
free field.

\textbf{(A3)} The system operators coupled to the environment have a
strength independent of the frequency ( this is due to a first Markov
approximation).

Assumptions (A2) and (A3) are necessary to obtain a quantum stochastic
differential equations and an idealized "white noise".

\textbf{(A4)} The measurement process is indirect by sensing the effect of
the system on the environment via a radiated field. If $b_{in}\left(
t\right) $ and $b_{out}\left( t\right) $ are an input field and an output
field respectively, the integrals $B_{in}\left( t\right)
=\int_{t_{0}}^{t}b_{in}\left( \tau \right) d\tau $ and $B_{out}\left(
t\right) =\int_{t_{0}}^{t}b_{out}\left( \tau \right) d\tau $ are interpreted
as a noise ( a quantum Wiener process) whenever the state of the field is
incoherent, e.g. a thermal equilibrium state or when the field is in vacuum.

\textbf{(A5)} We couple the open quantum system to $d$ measurement channels
(independent noise field inputs) via coupling operators $L_{i}$ ( for the
i-th channel). The indirect measurement is developed through a coupled
measurement channel playing the role of a quantum noise bath.

For a system of annihilation operators $\left\{ X_{k}:k=1,\ldots ,m\right\} $
and a system of creation operators \\
 $\left\{ X_{k+m}=X_{k}^{\dagger
}:k=1,\ldots ,m\right\} $, let $X_{-}$ be the stacking of annihilation
operators and $X_{+}$ the stacking of creation operators. We define the
state $\mathbb{X}$ in the Heisenberg picture as $\mathbb{X}=\left( 
\begin{array}{cc}
X_{-}^{T} & X_{+}^{T}%
\end{array}%
\right) ^{T}$. In a multiple-boson system the operators $X_{k}$ and $%
X_{k}^{\dagger }$ are not Hermitian and satisfy the canonical commutation
relations:%
\begin{eqnarray*}
\left[ X_{j},X_{k}\right] &=&\delta _{j+m,k}I \\
\left[ X_{j},X_{k}\right] &=&\left[ X_{j}^{\dagger },X_{k}^{\dagger }\right]
=0
\end{eqnarray*}%
where $\delta _{j,k}$ is the Dirac delta. Defining a matrix of state
commutation $\left[ \mathbb{X},\mathbb{X}\right] $ with $\left[ X_{i},X_{j}\right] $ as the
entry $\left( i,j\right) $, it is easy to verify that $\left[ \mathbb{X},\mathbb{X}\right]
=\left( \mathbb{S}\otimes I\right) $ where $\mathbb{S}$ stands for the
symplectic matrix 
\begin{equation*}
\mathbb{S}=\left( 
\begin{array}{cc}
0 & I \\ 
-I & 0%
\end{array}%
\right)
\end{equation*}%
Also we define the system Hamiltonian%
\begin{equation*}
H_{sys}=\frac{1}{2}\mathbb{X}^{\intercal }\left( R\otimes I\right) \mathbb{X}
\end{equation*}%
where $R=\left( 
\begin{array}{cc}
R_{11} & R_{12} \\ 
R_{21} & R_{22}%
\end{array}%
\right) \in \mathbb{R}^{2m\times 2m}$ is selected in such way that $H_{sys}$
is Hermitian. The adjoint of the system Hamiltonian operator is $H_{sys}^{\dagger }=\frac{1}{2}\mathbb{X}^{\intercal
}\left( \mathbb{J}R^{T}\mathbb{J}\otimes I\right) \mathbb{X}$ where $\mathbb{J}$ is the antidiagonal matrix%
\[
\mathbb{J}=\left(
\begin{array}
[c]{cc}%
0 & I\\
I & 0
\end{array}
\right)
\]
From this fact it follows that $%
R_{11}^{T}=R_{22}$, $R_{11}^{T}=R_{22}$, $R_{12}^{T}=R_{12}$ and $%
R_{21}^{T}=R_{21}$.

For each measurement channel we define a control action and these actions
are collected in a vector $u\in \mathbb{R}^{d}$. The control can be carried
out by the coupling of one or more tunable electromagnetic fields. We define
the controlled Hamiltonian $H\left( u\right) $ as%
\begin{eqnarray*}
H\left( u\right) &=&\frac{1}{2}\left( \mathbb{X}^{\intercal }\left( Ku\otimes
I\right) +\left( u^{T}K^{\dagger }\otimes I\right) \mathbb{X}\right) = \\
&=&\frac{1}{2}\left( u^{T}\left( K^{T}+K^{\dagger }\right) \otimes I\right) \mathbb{X}
\end{eqnarray*}%
where $K\in \mathbb{R}^{2m\times d}$ stands for a complex matrix of gains.
Let us write $K=\left( 
\begin{array}{c}
K_{-} \\ 
K_{+}%
\end{array}%
\right) $, for $H\left( u\right) $ to be Hermitian, $H\left( u\right) =$ $%
H\left( u\right) ^{\dagger }=\frac{1}{2}\left( u^{T}\left( K^{T}+K^{\dagger
}\right) \mathbb{J}\otimes I\right) \mathbb{X}$, and it is necessary that $\operatorname{Re}%
\left( K_{-}\right) =\operatorname{Re}\left( K_{+}\right) $.

We couple the open quantum system with internal Hamiltonian $H_{sys}$ and
controlled Hamiltonian $H\left( u\right) $ to $d$ measurement channels
(independent noise field inputs) via the vector operator $L=\left(
\Gamma \otimes I\right) \mathbb{X}$; this is an open quantum system with multiple
field channels where $\Gamma $ is an appropriate operator.

Then a quantum linear model in the state space representation based on
annihilators has the form%
\begin{equation}
d\mathbb{X}_{t}=\left( A\mathbb{X}_{t}+Bv_{t}\right) dt+dV_{t}  \label{EQ18}
\end{equation}%
\begin{equation*}
dY_{t}=\left( C\mathbb{X}_{t}+Dv_{t}\right) dt+dW_{t}
\end{equation*}%
where $V_{t}$ and $W_{t}$ represent quantum noises in the form of Wiener
process on a Fock space, instead of innovations of
the measurement process. Calling $\mathbb{A}$ and $\mathbb{A}^{\dagger}$to the stacking of operators
$A_{i}$ and $A_{i}^{\dagger}$, the noise increment in the state equation is
given by%
\[
dW\left(  t\right)  =\hbar\left(  \mathbb{S}\Gamma^{T}\mathbb{J}\otimes
I\right)  d\mathbb{A}^{\dagger}\left(  t\right)  +\hbar\left(  \mathbb{JS}%
\Gamma\otimes I\right)  d\mathbb{A}\left(  t\right)
\]

and the noise increment in the output is%
\[
dV\left(  t\right)  =\left(  d\mathbb{A}\left(  t\right)  +d\mathbb{A}%
^{\dagger}\left(  t\right)  \right)
\]

And the system matrices are given by
\[
A=\frac{\hbar}{2}\left(  \mathbb{S}\left(  R+R^{T}+\mathfrak{F}\left(
\mathbb{J}\Gamma^{\dagger}\Gamma\right)  \right)  \otimes I\right)
\]%
\[
B=\frac{1}{2}\left(  \mathbb{S}\left(  K+K^{\ast}\right)  \otimes I\right)
\]%
\[
C=\left(  \left(  \Gamma+\Gamma^{\ast}\right)  \otimes I\right)
\]%
\[
v\left(  t\right)  =\left(  u\left(  t\right)  \otimes I\right)
\]
with $\mathfrak{F}\left(  \mathbb{J}\Gamma^{\dagger}\Gamma\right)  =\Gamma
^{T}\Gamma^{\ast}\mathbb{J}+\mathbb{J}\Gamma^{\dagger}\Gamma$. In most
of the cases $\Gamma $ is real so that $\mathfrak{F}\left( \mathbb{J}\Gamma
^{\dagger }\Gamma \right) =0$.

The reader is referred to \cref{sec:Appendix} for more details in the derivation of \Cref{EQ18}.

\subsection{Hamilton-Jacobi-Bellman Equation in Expectation and Variance
Coordinates}

Let us define a complete set of coordinates describing the total probability
distribution given by $\rho $, \cite{parthasarathy2012introduction}:%
\begin{eqnarray*}
\hat{X}_{i} &=&\left\langle \rho ,X_{i}\right\rangle \\
\Sigma _{ij} &=&\left\langle \rho ,X_{i}X_{j}\right\rangle -\hat{X}_{i}\hat{X%
}_{j}
\end{eqnarray*}%
with dynamics, \cite{2005quant.ph..6018E},%
\begin{eqnarray*}
d\hat{X}_{t} &=&\left( A\hat{X}_{t}+Bu_{t}\right) dt+\tilde{K}_{t}d\tilde{Y}%
_{t} \\
\tilde{K}_{t} &=&\left( \Sigma C^{T}+M\right)
\end{eqnarray*}%
where $d\tilde{Y}_{t}$ is the innovation martingale and $M$ is a covariance
matrix of noise increments

Let us assume that we have achieved an optimal control action $u^{\ast
}\left( \cdot \right) $ in the interval $\left[ t+\Delta t,T\right] $.
Applying the optimality principle the problem is reduced to search for an
optimal solution $u\left( \cdot \right) $ in the interval $\left[ t,t+\Delta
t\right] $:%
\begin{equation*}
u\left( s\right) =\left\{ 
\begin{array}{cc}
u\left( s\right) & s\in \left[ t,t+\Delta t\right] \\ 
u^{\ast }\left( s\right) & s\in \left[ t+\Delta t,T\right]%
\end{array}%
\right.
\end{equation*}%
Under these conditions the cost-to-go function $S\left( t,\hat{X},\Sigma
\right) $ can be divided in two parts:%
\begin{gather}
S\left( t,\hat{X},\Sigma \right) =\min_{u_{\bullet }\left( \cdot \right) \in 
\mathcal{U}_{ad}\left[ t,T\right] }\mathbb{E}\left[ J_{\bullet }\left( t,%
\hat{X},\Sigma ;u_{\bullet }\left( \cdot \right) \right) \right] =
\label{EQ403} \\
=\min_{u_{\bullet }\left( \cdot \right) \in \mathcal{U}_{ad}\left[ t,T\right]
}\mathbb{E}\left[ \int_{t}^{t+\Delta t}\emph{C}\left( \tau ,u_{\bullet
}\left( \tau \right) ,\hat{X}_{\bullet }\left( \tau \right) ,\Sigma
_{\bullet }\left( \tau \right) \right) d\tau \right]  \notag \\
+S\left( t+\Delta t,\hat{X},\Sigma \right)  \notag
\end{gather}%
Note that $\hat{X}$ and $\Sigma $ are stochastic processes so that we can
apply the It\^{o}'s calculus:%
\begin{gather}
dS\left( t,\hat{X},\Sigma \right) =\frac{\partial S}{\partial t}%
+\left\langle \frac{\partial }{\partial \hat{X}},d\hat{X}\right\rangle
\left( S\right)  \label{EQ400} \\
+\frac{1}{2}\left\langle \frac{\partial }{\partial \hat{X}}\otimes \frac{%
\partial }{\partial \hat{X}},d\hat{X}\otimes d\hat{X}\right\rangle
+\left\langle \frac{\partial }{\partial \Sigma },d\Sigma \right\rangle
\left( S\right)  \notag
\end{gather}%
Let us observe that the term $d\Sigma \otimes d\Sigma $ is not included in $%
\left( \ref{EQ400}\right) $ since the differential equation for the
covariance matrix does not include uncertainty. As $dt\rightarrow 0$ the It%
\^{o}'s rules lead to $dt^{2}\rightarrow 0$, $d\tilde{Y}dt\rightarrow 0$,
and $\left( d\tilde{Y}\otimes d\tilde{Y}\right) \rightarrow dt$, resulting
in $d\hat{X}\otimes d\hat{X}=\left( \tilde{K}\otimes \tilde{K}\right) dt$,
and%
\begin{equation*}
dS\left( t,\hat{X},\Sigma \right) =\frac{\partial S}{\partial t}+\mathcal{D}%
S\left( t,\hat{X},\Sigma \right) dt+\left\langle \frac{\partial }{\partial 
\hat{X}},\tilde{K}d\tilde{Y}\right\rangle \left( S\right) dt
\end{equation*}%
where the stochastic differential operator $\mathcal{D}$ is now defined in
the following terms:%
\begin{gather}
\mathcal{D}:=\left\langle \frac{\partial \left( \cdot \right) }{\partial 
\hat{X}},A\hat{X}+Bu\right\rangle +\left\langle \left( \frac{\partial }{%
\partial \Sigma }\right) \left( \cdot \right) ,G\left( \Sigma \right)
\right\rangle  \label{EQ401} \\
+\frac{1}{2}\left\langle \left( \frac{\partial }{\partial \hat{X}}\otimes 
\frac{\partial }{\partial \hat{X}}\right) \left( \cdot \right) ,\left( 
\tilde{K}\otimes \tilde{K}\right) \right\rangle  \notag
\end{gather}%
with $G\left( \Sigma \right) =A\Sigma +\Sigma A^{T}-\tilde{K}_{t}^{T}\tilde{K%
}_{t}$. Writing $\left( \ref{EQ401}\right) $ in integral form yields%
\begin{gather}
S\left( t+\Delta t,\hat{X},\Sigma \right) =S\left( t,\hat{X},\Sigma \right)
\label{EQ402} \\
+\int_{t}^{t+\Delta t}\left( \frac{\partial S\left( \tau ,\hat{X},\Sigma
\right) }{\partial \tau }+\mathcal{D}S\left( \tau ,\hat{X},\Sigma \right)
\right) d\tau  \notag \\
+\int_{t}^{t+\Delta t}\left\langle \frac{\partial }{\partial \hat{X}},\tilde{%
K}d\tilde{Y}\right\rangle \left( S\left( \tau ,\hat{X},\Sigma \right)
\right) d\tau  \notag
\end{gather}%
Folding $\left( \ref{EQ402}\right) $ into $\left( \ref{EQ403}\right) $,%
\begin{gather}
S\left( t,\hat{X},\Sigma \right) =\min_{u_{\bullet }\left( \cdot \right) \in 
\mathcal{U}_{ad}\left[ t,T\right] }\mathbb{E[}S\left( t,\hat{X},\Sigma
\right)  \label{EQ404} \\
+\int_{t}^{t+\Delta t}\emph{C}\left( \tau ,u_{\bullet }\left( \tau \right) ,%
\hat{X}_{\bullet }\left( \tau \right) ,\Sigma _{\bullet }\left( \tau \right)
\right) d\tau  \notag \\
+\int_{t}^{t+\Delta t}\left( \frac{\partial S\left( \tau ,\hat{X},\Sigma
\right) }{\partial \tau }+\mathcal{D}S\left( \tau ,\hat{X},\Sigma \right)
\right) d\tau ]  \notag
\end{gather}%
where we have account for 
\begin{equation*}
\mathbb{E}\left[ \int_{t}^{t+\Delta t}\left\langle \frac{\partial }{\partial 
\hat{X}},\tilde{K}d\tilde{Y}\right\rangle \left( S\left( \tau ,\hat{X}%
,\Sigma \right) \right) d\tau \right] =0
\end{equation*}
for being $\tilde{Y}$ a Gaussian process.

Now we can take expectations on both sides of $\left( \ref{EQ404}\right) $
and put $S\left( t,\hat{X},\Sigma \right) $ out of the minimization since it
does not depend on $u$:%
\begin{gather}
0=\min_{u_{\bullet }\left( \cdot \right) \in \mathcal{U}_{ad}\left[ t,T%
\right] }\mathbb{E[}\int_{t}^{t+\Delta t}(\emph{C}\left( \tau ,u_{\bullet
}\left( \tau \right) ,\hat{X}_{\bullet }\left( \tau \right) ,\Sigma
_{\bullet }\left( \tau \right) \right) \\
+\frac{\partial S\left( \tau ,\hat{X},\Sigma \right) }{\partial \tau }+%
\mathcal{D}S\left( \tau ,\hat{X},\Sigma \right) )d\tau ]  \notag
\end{gather}%
In order to satisfy this expression the integrand should be zero and
interchanging the minimum operation with the integral the HJB equation in
the Heisenberg picture is finally derived:%
\begin{equation*}
\begin{split}
\min_{u\left( \cdot \right) \in \mathcal{U}_{ad}\left[ t,T\right] }\left\{ 
\emph{C}\left( t,u,\hat{X},\Sigma \right) + 
\frac{\partial S\left( t,\hat{X}
,\Sigma \right) }{\partial t}+ \right. \\
\left. \mathcal{D}S\left( t,\hat{X},\Sigma \right)
\right\} =0
\end{split}
\end{equation*}%
or equivalently%
\begin{equation*}
-\frac{\partial S\left( t,\hat{X},\Sigma \right) }{\partial t}=\min_{u\left(
\cdot \right) \in \mathcal{U}_{ad}\left[ t,T\right] }\mathcal{H}\left( t,u,%
\hat{X},\Sigma \right)
\end{equation*}%
where%
\begin{gather}
\mathcal{H}\left( t,u,\hat{X},\frac{\partial S}{\partial \hat{X}},\left( 
\frac{\partial }{\partial \hat{X}}\otimes \frac{\partial }{\partial \hat{X}}%
\right) \left( S\right) ,\Sigma ,\frac{\partial S}{\partial \Sigma }\right) =
\label{EQ405} \\
\emph{C}\left( t,u,\hat{X},\Sigma \right) +\mathcal{D}S\left( t,\hat{X}%
,\Sigma \right)  \notag
\end{gather}%
Henceforth, the control action $u\left( \cdot \right) $ minimizing $S$ can
be found in terms of $t$, $\hat{X}$, $\frac{\partial S}{\partial \hat{X}}$, $%
\left( \frac{\partial }{\partial \hat{X}}\otimes \frac{\partial }{\partial 
\hat{X}}\right) \left( S\right) $, $\Sigma $, and $\frac{\partial S}{%
\partial \Sigma }$.

In the derivation of the HJB equation in the Heisenberg picture the value
function $S$ was assumed to be twice differentiable. Now we provide
necessary and sufficient conditions for $S$ to be $C^{2}$ with
respect to the states $\hat{X}$ and $\Sigma$. Under these conditions we
firstly prove that the solutions of the stochastic HJB are twice
differentiable in the interior of the state space. Secondly, the value
function is proved to be a solution of the HJB.

The stochastic HJB can be rewritten as follows:%
\begin{equation}
-\frac{1}{2}\left\langle \left(  \frac{\partial}{\partial\hat{X}}\otimes
\frac{\partial}{\partial\hat{X}}\right)  \left(  S\right)  ,\left(  \tilde
{K}\otimes\tilde{K}\right)  \right\rangle =\mathcal{G}\left(  \hat{X}%
,\Sigma,\frac{\partial S}{\partial\hat{X}},\frac{\partial S}{\partial\Sigma
},\frac{\partial S}{\partial t}\right)  \label{EQ500}%
\end{equation}
where%
\begin{align*}
\mathcal{G}\left(  \hat{X},\Sigma,\frac{\partial S}{\partial\hat{X}}%
,\frac{\partial S}{\partial\Sigma},\frac{\partial S}{\partial t}\right)    &
=\min_{u\in U}g\left(  u,\hat{X},\Sigma,\frac{\partial S}{\partial\hat{X}%
},\frac{\partial S}{\partial\Sigma},\frac{\partial S}{\partial t}\right)  \\
g\left(  u,\hat{X},\Sigma,p_{X},p_{\Sigma},r\right)    & =r+\emph{C}\left(
t,u,\hat{X},\Sigma\right)  +\left\langle p_{X},A\hat{X}+Bu\right\rangle
+\left\langle p_{\Sigma},G\left(  \Sigma\right)  \right\rangle
\end{align*}

The following two assumptions guarantee the existence and uniqueness of the
solutions of the stochastic differential equation.

\begin{description}
\item[Assumption (H1):] The cost $\emph{C}$ is uniformly continuous in $u$ and
Lipschitz in $\hat{X}$, $\Sigma$, on any compact set of $\mathbb{R}^{n}%
\times\mathbb{R}^{n\times n}$.

\item[Assumption (H2):] $G\left(  \Sigma\right)  $ is Lipschitz in $\Sigma$.
\end{description}

\subsection{Pontryagin Principle in Coordinates $\hat{X}$ and $\Sigma $}

On the basis of the HJB formulation in the Heisenberg picture we can derive
the Pontryagin's maximum principle. To this end we previously define the
adjoint variables for the quantum optimal control problem:%
\begin{eqnarray}
p_{X} &=&\frac{\partial S\left( t,\hat{X},\Sigma \right) }{\partial \hat{X}}
\label{EQ406} \\
q_{X} &=&\left( \frac{\partial }{\partial \hat{X}}\otimes \frac{\partial }{%
\partial \hat{X}}\right) \left( S\left( t,\hat{X},\Sigma \right) \right) 
\notag \\
p_{\Sigma } &=&\frac{\partial S\left( t,\hat{X},\Sigma \right) }{\partial
\Sigma }  \notag
\end{eqnarray}%
Note that the covariance matrix $\Sigma $ has a
deterministic dynamics ( without Gaussian noise) so that $q_{\Sigma }=0$.
Also we define the Hamiltonian function $\mathcal{H}$ in the Heisenberg
picture as%
\begin{gather*}
\mathcal{H}\left( t,u,\hat{X},p_{X},q_{X},\Sigma ,p_{\Sigma }\right) =\emph{C%
}\left( t,u,\hat{X},\Sigma \right) \\
+\left( \left\langle p_{X},A\hat{X}+Bu\right\rangle +\frac{1}{2}\left\langle
q_{X},\left( \tilde{K}\otimes \tilde{K}\right) \right\rangle +\left\langle
p_{\Sigma },G\left( \Sigma \right) \right\rangle \right)
\end{gather*}%
With these definitions the quantum version of $\left( \ref{EQ406}\right) $
is as follows%
\begin{equation*}
-\frac{\partial S\left( t,\hat{X},\Sigma \right) }{\partial t}=\min_{u\left(
\cdot \right) \in \mathcal{U}\left[ t,T\right] }\mathcal{H}\left( t,u,\hat{X}%
,p_{X},q_{X},\Sigma ,p_{\Sigma }\right)
\end{equation*}%
At this point it is assumed that there exists a unique minimizing control
law \\
$u^{\ast }\left( t,\hat{X},p_{X},q_{X},\Sigma ,p_{\Sigma }\right)$ such
that%
\begin{equation*}
\mathcal{H}\left( t,u^{\ast },\hat{X},p_{X},q_{X},\Sigma ,p_{\Sigma }\right)
=\inf_{u\left( \cdot \right) \in \mathcal{U}}\mathcal{H}\left( t,\hat{X}%
,p_{X},q_{X},\Sigma ,p_{\Sigma }\right)
\end{equation*}%
The dynamics of the expectation $\hat{X}$ can be derived directly from the
definition of $\mathcal{H}$:%
\begin{equation*}
d\hat{X}=\frac{\partial \mathcal{H}}{\partial p_{X}}dt+\tilde{K}d\tilde{Y}
\end{equation*}%
On the other hand the dynamics of the adjoint operators $p_{X}\left( t,\hat{X%
},\Sigma \right)$ and \\$p_{\Sigma }\left( t,\hat{X},\Sigma \right)$ can be
determined by resorting to the It\^{o}'s Lemma:%
\begin{eqnarray*}
dp_{X} &=&\frac{\partial p_{X}}{\partial t}dt+\mathcal{D}p_{X}+\left\langle 
\frac{\partial }{\partial \hat{X}}\left( \cdot \right) ,\tilde{K}d\tilde{Y}\right\rangle
\left( p_{X}\right) \\
dp_{\Sigma } &=&\frac{\partial p_{\Sigma }}{\partial t}dt+\mathcal{D}%
p_{\Sigma }
\end{eqnarray*}%
In view of $\left( \ref{EQ406}\right) $ it follows that 
\begin{eqnarray*}
\frac{\partial p_{X}}{\partial t} &=&\frac{\partial }{\partial \hat{X}}\frac{%
\partial S\left( t,\hat{X},\Sigma \right) }{\partial t} \\
\frac{\partial p_{\Sigma }}{\partial t} &=&\frac{\partial }{\partial \Sigma }%
\frac{\partial S\left( t,\hat{X},\Sigma \right) }{\partial t}
\end{eqnarray*}%
and from the chain rule of the differentiation:%
\begin{eqnarray*}
-\frac{\partial p_{X}}{\partial t} &=&\frac{\partial \mathcal{H}}{\partial 
\hat{X}}+\mathcal{D}p_{X} \\
-\frac{\partial p_{\Sigma }}{\partial t} &=&\frac{\partial \mathcal{H}}{%
\partial \Sigma }+\mathcal{D}p_{\Sigma }
\end{eqnarray*}%
Henceforth,%
\begin{eqnarray*}
dp_{X} &=&-\frac{\partial \mathcal{H}}{\partial \hat{X}}dt+\left\langle \frac{%
\partial }{\partial \hat{X}}\left( \cdot \right) ,\tilde{K}d\tilde{Y}\right\rangle \left(
p_{X}\right) \\
dp_{\Sigma } &=&-\frac{\partial \mathcal{H}}{\partial \Sigma } dt
\end{eqnarray*}%
We can write the system of forward-backward quantum differential equations
for the Quantum Pontryagin principle in the Heisenberg picture:%
\begin{equation}
\left\{ 
\begin{array}{c}
d\hat{X}^{\ast }=\frac{\partial \mathcal{H}^{\ast }}{\partial p_{X}}%
dt+\tilde{K}d\tilde{Y} \\ 
dp_{X}^{\ast }=-\frac{\partial \mathcal{H}^{\ast }}{\partial \hat{X}}%
dt+\left\langle \frac{\partial }{\partial \hat{X}}\left( \cdot \right)
,\tilde{K}d\tilde{Y}\right\rangle \left( p_{X}^{\ast }\right) \\ 
dp_{\Sigma }^{\ast }=-\frac{\partial \mathcal{H}^{\ast }}{\partial \Sigma }
\\ 
\hat{X}^{\ast }\left( 0\right) =\hat{X}_{0} \\ 
p_{X}^{\ast }\left( T\right) =\frac{\partial \mathcal{S}\left( T,\hat{X}%
^{\ast }\left( T\right) ,\Sigma ^{\ast }\left( T\right) \right) }{\partial 
\hat{X}} \\ 
p_{\Sigma }^{\ast }\left( T\right) =\frac{\partial \mathcal{S}\left( T,\hat{X%
}^{\ast }\left( T\right) ,\Sigma ^{\ast }\left( T\right) \right) }{\partial
\Sigma }%
\end{array}%
\right.  \label{EQ410}
\end{equation}%
As a matter of fact $\Sigma $ is 
governed by a deterministic Riccati differential equation so it is not
necessary to compute its costate.

\subsection{Stochastic LQG control from the Quantum Pontryagin Principle}

In this section we obtain an LQG control as a direct consequence of the principle of the Pontryagin maximum. The main difference with respect to other methods that appear in the literature is that these methods are based on the Bellman optimality principle, and this is where the main difference is found, \cite{Doherty1999}.

The forward-backward quantum differential equations in $\left( \ref{EQ410}%
\right) $ cannot be solved in closed form so it is mandatory resorting to
numerical solutions. As an attempt to avoid this numerical computation in
this section a LQG scheme will be derived from the quantum PMP. Let us
consider the following dynamics and cost functional:

\begin{gather*}
d\hat{X}=\left( A\hat{X}+Bu\right) dt+\tilde{K}d\tilde{Y} \\
\hat{X}\left( 0\right) =\hat{X}_{0} \\
J\left( u\right) =E[\hat{X}^{T}\left( T\right) F\hat{X}\left( T\right) \\
+\frac{1}{2}\int_{0}^{T}\left( \hat{X}^{T}\left( t\right) Q\left( t\right) 
\hat{X}\left( t\right) +u^{T}\left( t\right) R\left( t\right) u\left(
t\right) \right) dt]
\end{gather*}%
with $F\succeq 0$, $Q\left( t\right) ,R\left( t\right) \succ 0$. The
Hamiltonian function $\mathcal{H}$ is defined as%
\begin{gather*}
\mathcal{H}\left( t,\hat{X},u,p_{X},q_{X}\right) =\frac{1}{2}\left( \hat{X}%
^{T}Q\hat{X}+u^{T}Ru\right) \\
+\left\langle p_{X},A\hat{X}+Bu\right\rangle +\frac{1}{2}\left\langle
q_{X},\left( \tilde{K}\otimes \tilde{K}\right) \right\rangle
\end{gather*}%
and the Pontryagin's necessary conditions are:%
\begin{equation*}
\left\{ 
\begin{array}{c}
d\hat{X}^{\ast }=\left( A\hat{X}^{\ast }+Bu^{\ast }\right) dt+\tilde{K}d\tilde{Y} \\ 
dp_{X}^{\ast }=-\left( Q\hat{X}^{\ast }+A^{T}p_{X}^{\ast }\right)
dt+\left\langle \frac{\partial }{\partial \hat{X}}\left( \cdot \right)
,\tilde{K}d\tilde{Y}\right\rangle \left( p_{X}^{\ast }\right) \\ 
\hat{X}^{\ast }\left( 0\right) =\hat{X}_{0} \\ 
p_{X}^{\ast }\left( T\right) =F\hat{X}^{\ast }\left( T\right)%
\end{array}%
\right.
\end{equation*}%
From the inequality $\mathcal{H}\left( t,u^{\ast },\hat{X}^{\ast
},p_{X}^{\ast },q_{X}^{\ast }\right) \leq \mathcal{H}\left( t,u,\hat{X}%
^{\ast },p_{X}^{\ast },q_{X}^{\ast }\right) $ we can obtain an optimal
control%
\begin{eqnarray*}
u^{\ast }\left( t\right) &=&\arg \min_{u}\frac{1}{2}\left( u^{T}\left(
t\right) R\left( t\right) u\left( t\right) +u^{T}\left( t\right)
B^{T}p_{X}^{\ast }\left( t\right) \right) = \\
&=&-R^{-1}\left( t\right) B^{T}\left( t\right) p_{X}^{\ast }\left( t\right)
\end{eqnarray*}%
Folding this control minimizing $\mathcal{H}$ into the differential
equations for $\hat{X}^{\ast }$ and $p^{\ast }$ yields:%
\begin{eqnarray*}
d\hat{X}^{\ast } &=&\left( A\hat{X}^{\ast }-BR^{-1}\left( t\right)
B^{T}\left( t\right) p_{X}^{\ast }\left( t\right) \right) dt+\tilde{K}d\tilde{Y} \\
\hat{X}^{\ast }\left( 0\right) &=&\hat{X}_{0}
\end{eqnarray*}%
Given that $p_{X}^{\ast }$ linearly depends on $\hat{X}^{\ast }$ we can use
the ansatz $p_{X}^{\ast }\left( t\right) =K\left( t\right) \hat{X}+\varphi
\left( t\right) $, and applying the rules of the stochastic differential
calculus along with the fact that $\frac{\partial p_{X}^{\ast }\left(
t\right) }{\partial \hat{X}}=\left\langle \frac{\partial }{\partial \hat{X}}%
\left( \cdot \right) ,\tilde{K}d\tilde{Y}\right\rangle \left( p_{X}^{\ast }\right) =K\left(
t\right) $ it is concluded that%
\begin{gather}
dp_{X}^{\ast }=(K^{\prime }\left( t\right) \hat{X}^{\ast }+\varphi ^{\prime
}\left( t\right)  \label{EQ407} \\
+K\left( t\right) \left[ A\hat{X}^{\ast }-BR^{-1}\left( t\right) B^{T}\left(
t\right) \left( K\left( t\right) \hat{X}^{\ast }+\varphi \left( t\right)
\right) \right] )dt  \notag \\
+\left\langle \frac{\partial }{\partial \hat{X}}\left( \cdot \right)
,\tilde{K}d\tilde{Y}\right\rangle \left( p_{X}^{\ast }\right)  \notag
\end{gather}%
On the other hand, from the backward-forward system,%
\begin{gather}
dp_{X}^{\ast }=-\left( Q\hat{X}^{\ast }+A^{T}\left( K\left( t\right) \hat{X}%
^{\ast }+\varphi \left( t\right) \right) \right) dt  \label{EQ408} \\
+\left\langle \frac{\partial }{\partial \hat{X}}\left( \cdot \right)
,\tilde{K}d\tilde{Y}\right\rangle \left( p_{X}^{\ast }\right)  \notag
\end{gather}%
\bigskip Matching $\left( \ref{EQ407}\right) $ and $\left( \ref{EQ408}%
\right) $,%
\begin{gather}
K^{\prime }\left( t\right) \hat{X}^{\ast }+\varphi ^{\prime }\left( t\right)
+K\left( t\right) [A\hat{X}^{\ast }  \label{EQ409} \\
-BR^{-1}\left( t\right) B^{T}\left( t\right) \left( K\left( t\right) \hat{X}%
^{\ast }+\varphi \left( t\right) \right) ]  \notag \\
=-\left( Q\left( t\right) \hat{X}^{\ast }+A^{T}\left( K\left( t\right) \hat{X%
}^{\ast }+\varphi \left( t\right) \right) \right)  \notag
\end{gather}%
This finally leads to the differentials equations for $K\left( t\right) $
and $\varphi $:%
\begin{gather*}
K^{\prime }\left( t\right) =-K\left( t\right) A-A^{T}K\left( t\right) \\
+K\left( t\right) BR^{-1}\left( t\right) B^{T}\left( t\right) K\left(
t\right) -Q\left( t\right) \\
K\left( T\right) =F \\
\varphi ^{\prime }\left( t\right) =-\left[ A-BR^{-1}\left( t\right)
B^{T}K\left( t\right) \right] ^{T}\varphi \left( t\right) \\
\varphi \left( T\right) =0
\end{gather*}

\section{Example: Controlled Harmonic Oscillator}
\label{sec:Example}

We illustrate the above ideas for the case of a dissipative quantum harmonic
oscillator. Specifically we will consider a single measurement channel with an
antihermitian measurement operator. Let us define the state space vector
operator as consisting of a creation operator (or raising operator) $a$ and an
annihilation operator $a^{\dagger}$ (or lowering operator):
\[
\mathbb{X}=\left(
\begin{array}
[c]{c}%
a\\
a^{\dagger}%
\end{array}
\right)
\]
where the creation operator has been taken from \cite{Jacobs06}:
\[
a=\frac{1}{\sqrt{2}x_{0}}X_{Q}+\mathbf{i}\frac{x_{0}}{\sqrt{2}\hbar}X_{P}%
\]
with $x_{0}=\sqrt{\frac{\hbar}{m\omega_{0}}}$ and $\omega_{0}$ being the
resonance frequency. We can recover the generalized position $X_{Q}$ and
momentum $X_{P}$ from the creation and annihilation operators:%

\begin{align*}
X_{Q}  & =\frac{x_{0}}{\sqrt{2}}\left(  a+a^{\dagger}\right) \\
X_{P}  & =\frac{\sqrt{2}\hbar}{x_{0}}\frac{a-a^{\dagger}}{2\mathbf{i}}%
\end{align*}
The system Hamiltonian is defined as $H_{sys}\left(  X_{Q},X_{P}\right)
=\frac{X_{P}^{2}}{2m}+\frac{1}{2}m\omega_{0}^{2}X_{Q}^{2}$ and in
creation-annihilation coordinates as $H_{sys}\left(  a,a^{\dagger}\right)
=\frac{1}{2}\mathbb{X}^{\intercal}\left(  P\otimes I\right)  \mathbb{X}$ with
\[
P=\frac{\hbar^{2}}{4mx_{0}^{2}}\left(
\begin{array}
[c]{cc}%
-1 & 1\\
1 & -1
\end{array}
\right)  +\frac{m\omega_{0}^{2}}{4}\left(
\begin{array}
[c]{cc}%
1 & 1\\
1 & 1
\end{array}
\right)
\]
The controlled Hamiltonian depends on the operators creation and annihilation
as $H\left(  u\right)  =-u\frac{a+a^{\dagger}}{\sqrt{2}}$. In general we can
transform the system description into generalized coordinates of position and
momentum $\mathbb{Z=}\left(  \mathbb{X}_{Q},\mathbb{X}_{P}\right)  $ through
the transformation $\mathbb{Z}=\mathbb{TX}$ with:%

\begin{align*}
\mathbb{T}  & =T\otimes I\\
T  & =\left(
\begin{array}
[c]{cc}%
0 & \frac{1}{2}D_{1}\\
\frac{1}{2\mathbf{i}}D_{2} & 0
\end{array}
\right)  \left(  I-\mathbb{S}\right)
\end{align*}
where $D_{1}$ and $D_{2}$ are appropariate diagonal matrices. Note that $T$ is
a complex matrix:%
\[
T=\left(
\begin{array}
[c]{cc}%
0 & \frac{1}{2}x_{0}\sqrt{2}\\
\frac{1}{2\mathbf{i}}\frac{\sqrt{2}\hbar}{x_{0}} & 0
\end{array}
\right)  \left(
\begin{array}
[c]{cc}%
1 & -1\\
1 & 1
\end{array}
\right)  =\left(
\begin{array}
[c]{cc}%
\frac{1}{2}\sqrt{2}x_{0} & \frac{1}{2}\sqrt{2}x_{0}\\
-\frac{1}{2}i\sqrt{2}\frac{\hbar}{x_{0}} & \frac{1}{2}i\sqrt{2}\frac{\hbar
}{x_{0}}%
\end{array}
\right)
\]
The open quantum system is coupled to a single measurement channel for the
position via the operator $L$:
\begin{equation}
L=\left(  \Gamma\otimes I\right)  \mathbb{X}\text{ where }\Gamma=\left(
\begin{array}
[c]{cc}%
1 & 0
\end{array}
\right)  T=\frac{1}{2}\sqrt{2}x_{0}\left(
\begin{array}
[c]{cc}%
1 & 1
\end{array}
\right) \label{EQ515}%
\end{equation}
where $\Gamma$ is such that $\Gamma^{T}\Gamma^{\ast}\mathbb{J}+\mathbb{J}%
\Gamma^{\dagger}\Gamma$ is a matrix with entries $x_{0}^{2}$ and then%
\begin{align*}
A  & =\mathbb{S}\left(  \hbar R+\frac{\hbar}{2}x_{0}^{2}\left(
\begin{array}
[c]{cc}%
1 & 1\\
1 & 1
\end{array}
\right)  \right) \\
C  & =\left(  \Gamma+\Gamma^{\ast}\right)  =\sqrt{2}x_{0}\left(
\begin{array}
[c]{cc}%
1 & 1
\end{array}
\right)
\end{align*}
For a Hermitian matrix $K\in\mathbb{R}^{2\times2}$, $B$ is $\mathbb{S}K$; for
instance if $K$ is the diagonal matrix $diag\left(  K_{1}K_{2}\right)  $,
$B=diag\left(  K_{2},-K_{1}\right)  $. With the triplet $\left(  A,B,C\right)
$ the state-space equations is a forward differential equation%
\begin{align*}
d\hat{X}\left(  t\right)   & =\left(  A\hat{X}\left(  t\right)  +Bu\left(
t\right)  \right)  dt+\tilde{K}\left(  t\right)  d\tilde{Y}\left(  t\right) \\
\hat{X}\left(  0\right)   & =X_{0}%
\end{align*}
where the innovation process $\tilde{Y}\left(  t\right)  $ describes the gain
of information due to measurement of $Y\left(  t\right)  $. We define the cost
functional in coordinates $\hat{X}$ as
\[
J\left(  u\right)  =\frac{1}{2}\mathbb{E}\left[  \int_{0}^{T}\left(  \hat
{X}^{T}\left(  t\right)  S\hat{X}\left(  t\right)  +u^{T}\left(  t\right)
Ru\left(  t\right)  \right)  dt\right]
\]
The quantum Pontryagin principle results into the backward differential
equation%
\[%
\begin{array}
[c]{c}%
dp_{X}^{\ast}\left(  t\right)  =-\left(  S\hat{X}\left(  t\right)  +A^{T}%
p_{X}^{\ast}\left(  t\right)  \right)  dt+\left\langle \frac{\partial
}{\partial\hat{X}}\left(  \cdot\right)  ,\tilde{K}\left(  t\right)  d\tilde
{Y}\left(  t\right)  \right\rangle \left(  p_{X}^{\ast}\left(  t\right)
\right) \\
p_{X}^{\ast}\left(  T\right)  =0
\end{array}
\]
In general deriving the solution of a stochastic optimal control law is a
difficult task, since the nonlinear forward-backward stochastic differential equations
 are hardly ever solvable. However
exploiting the special structure of the LQG problem allows us to find an
explicit control law. To illustrate this point we rewrite the quantum harmonic
oscillator in generalized position and momentum coordinates from the creation
and annihilation operators in the model. For the sake of clarity and
concreteness in the exposition we borrow directly the equations of the quantum
harmonic oscillator from \cite{Jacobs06}. The equations have been
adapted to the formalism of Belavkin to show clearly the innovation process:%

\begin{align*}
A  & =\left(
\begin{array}
[c]{cc}%
-\frac{\gamma}{2} & \frac{1}{m}\\
-m\omega_{0}^{2} & -\frac{\gamma}{2}%
\end{array}
\right)  \text{, }B=\left(
\begin{array}
[c]{c}%
0\\
1
\end{array}
\right)  \text{, }C=\left(
\begin{array}
[c]{cc}%
\sqrt{2m\gamma\eta\frac{\omega_{0}}{\hbar}} & 0\\
0 & 0
\end{array}
\right) \\
M  & =\left(
\begin{array}
[c]{cc}%
-\sqrt{\frac{\gamma\eta\hbar}{2m\omega_{0}}} & 0\\
0 & 0
\end{array}
\right)  \text{, }N=\left(
\begin{array}
[c]{cc}%
\frac{1}{2m}\frac{\gamma}{\omega_{0}}\hbar & 0\\
0 & \frac{1}{2}m\gamma\omega_{0}\hbar
\end{array}
\right) \\
\dot{\Sigma}  & =A\Sigma+\Sigma A^{T}+N-\left(  \Sigma C^{T}+M\right)  \left(
\Sigma C^{T}+M\right)  ^{T}\\
\tilde{K}\left(  t\right)   & =\Sigma C^{T}+M\\
d\tilde{Y}\left(  t\right)   & =dY\left(  t\right)  -C\hat{X}\left(  t\right)
dt
\end{align*}
where $\gamma$ denotes the spontaneous emission rate, $\eta$ the efficiency of
the detector, $\omega_{0}$ the resonance frequency and $m$ the mass of the particle.

\begin{figure}[h]
  \centering
	\includegraphics[ height=9cm, width=13cm]{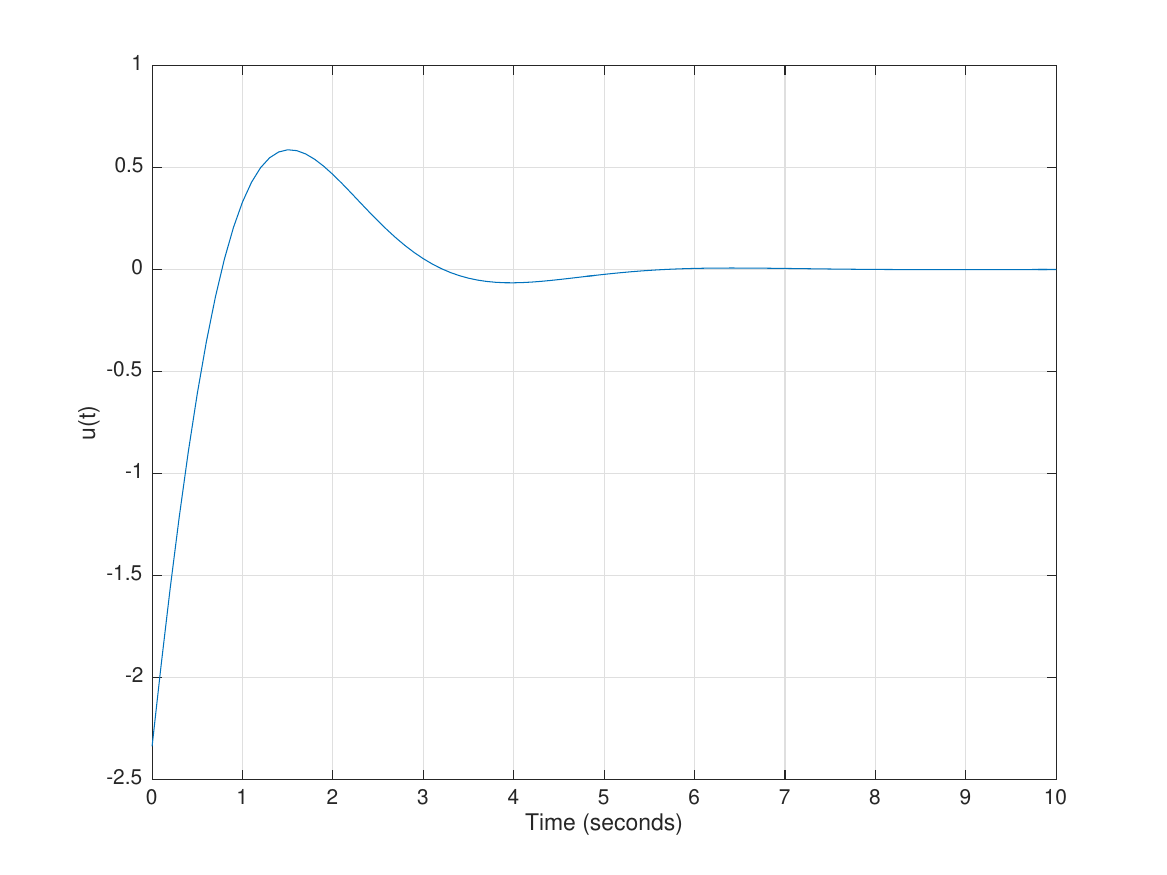}
  \caption{Precomputed Control signal for LQG control.}
  \label{FIG1}
\end{figure}

We use the ansatz $p_{X}\left(  t\right)  =K\left(  t\right)  \hat{X}%
+\varphi\left(  t\right)  $ where $K\left(  t\right)  $ is governed by the
following differential matrix Riccati differential equation:%
\begin{align}
\dot{K}\left(  t\right)   & =-K\left(  t\right)  A-A^{T}K\left(  t\right)
+K\left(  t\right)  BR^{-1}B^{T}K\left(  t\right)  +Q\label{EQ801}\\
K\left(  T\right)   & =F\nonumber
\end{align}
And $\varphi\left(  t\right)  $ has a dynamics driven by a backward
differential equation:%
\begin{align}
\dot{\varphi}\left(  t\right)   & =-\left[  A-BR^{-1}B^{T}K\left(  t\right)
\right]  ^{T}\varphi\left(  t\right) \label{EQ800}\\
\varphi\left(  T\right)   & =0\nonumber
\end{align}
This model has been simulated in MATLAB with both the Financial Toolbox
(Euler-Maruyama integration method) and the Toolbox SDETool for the numerical
solution of stochastic differential equations (SDEs) developed by Andrew
Horchler (, and the Milstein integration method). The parameters of the
quantum harmonic oscillator are $\gamma=0.1$, $\eta=0.1$, $\hbar=1$,
$m=\frac{1}{2}$, and $\omega_{0}=1$ rad/seg. The cost functional is built with
the matrices $R=Q=20I_{2}$ \thinspace where $I_{2}$ stands for the $2\times2$
identity matrix. The terminal time is chosen as $T=10$ sec. and the
integration step was chosen as $dt=0.1$ sec.

The initial conditions for the problem were selected so as to satisfy the
Heisenberg uncertainty principle,
\begin{align*}
\hat{X}\left(  0\right)    & =\left(  1,1\right)  \\
\Sigma\left(  0\right)    & =\left(
\begin{array}
[c]{cc}%
1 & 1\\
1 & \frac{\hbar^{2}}{2}%
\end{array}
\right)
\end{align*}
The action control $u\left(  t\right)  $ is precomputed according to the
ansatz via the expression $-R^{-1}B^{T}\left(  K\left(  t\right)  \hat
{X}+\varphi\left(  t\right)  \right)  $. Since the backward differential
equations for $K\left(  t\right)  $ and $\varphi\left(  t\right)  $ are hard
to be analytically solved we have decided on numerically solving them through
the Runge-Kutta (4,5) method with integration step $dt$ and then interpolating
the result with splines. The precomputed control signal is shown in \Cref{FIG1} where the action control decreases as the system gains information. The
asymptotic behaviour is such that $\lim_{t\rightarrow\infty}u\left(  t\right)
=0$ since the uncertainty diminishes to zero as time goes to infinity.

\begin{figure}[h]
  \centering
  \subfloat[Uncontrolled QHO.]{\label{fig:a}\includegraphics[ height=6cm, width=6.2cm]{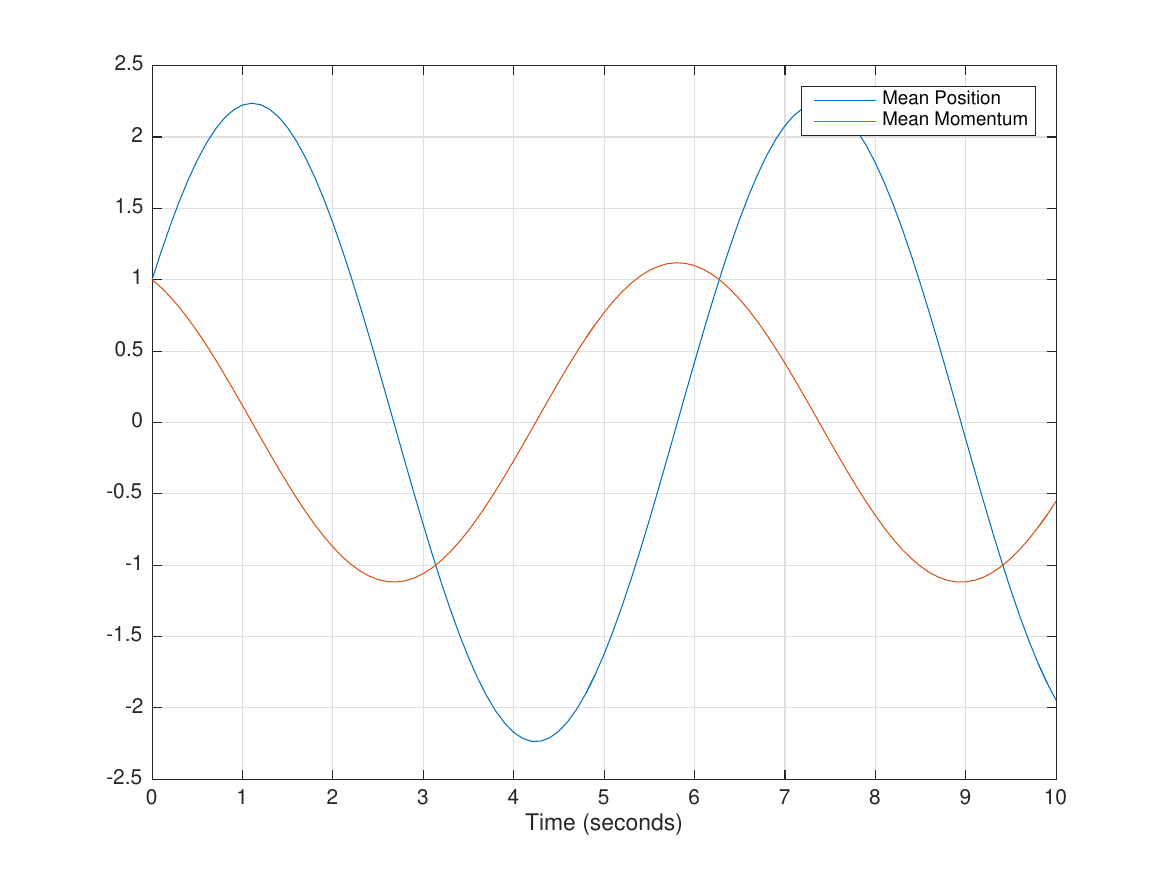} }
  \subfloat[Controlled QHO with a LQG controller]{\label{fig:b}\includegraphics[ height=6cm, width=6.2cm]{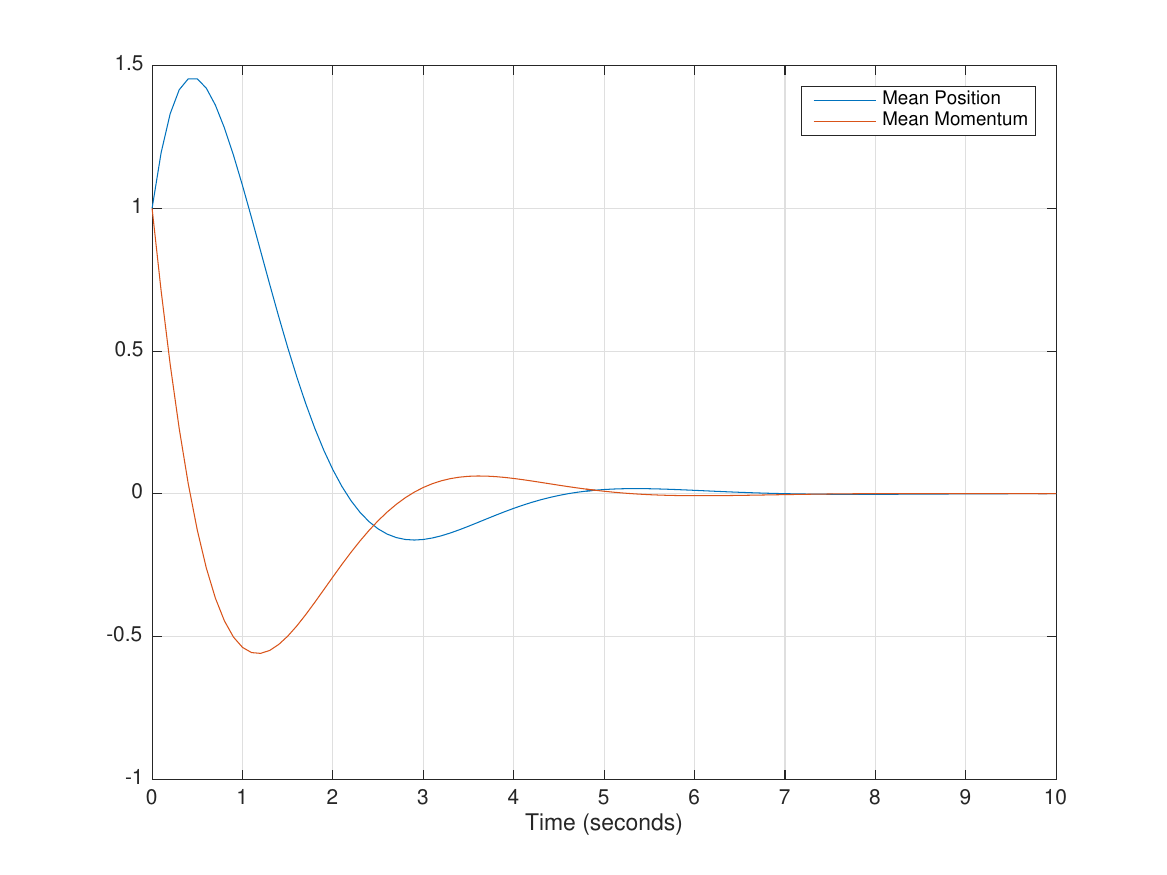}}
  \caption{Mean position and mean momentum for the Quantum Harmonic Oscillator.}
  \label{fig:Simfig}
\end{figure}

The position and the momentum of the particle are plotted both for the
uncontrolled case in \Cref{fig:a} and for the LQG control in \Cref{fig:b}.

\section{Conclusions}
\label{sec:Conclusions}

A quantum maximum principle has been addressed for \\continuous-time
measurements. This has
been derived from the Hamilton-Jacobi-Bellman equation in the Schr\"{o}%
dinger picture. Then the scheme has been extended to the Heisenberg picture
in statistical moment coordinates. Since a stochastic Pontryagin principle
requires a numerical solution to solve the equations we have derived a LQG
scheme which is more suitable for control purposes.

The Pontryagin principle tapes its roots in the method of Lagrange
multipliers applied in constrained optimization. A constraint on the state
essentially introduces infinitely many additional constraints compared with
a deterministic state constrained control problem; PMP approach allows to
transform an infinite dimensional optimization problem (the search over a
set of functions) into a finite dimensional optimization problem (the search
over a set of parameters). In the future it should be interesting to explore
QOC problems under state constraints.

\appendix
\section{Quantum Linear Model}
\label{sec:Appendix}

Let $H_{sys}$ be the free Hamiltonian of the system, and let $L_{k}$ and
$S_{k}$ be bounded system operators specifying the coupling and the scattering
of the system to the k-th measurement channel respectively. The
operator-valued process $U\left(  t\right)  $ representing the joint evolution
of the composite quantum system has a dynamics governed by a quantum
stochastic differential equation (QSDE) of the form:%
\begin{align}
dU\left(  t\right)   & =\left(  -\frac{\mathbf{i}}{\hbar}\left(
H_{sys}+H\left(  u\right)  \right)  +\frac{1}{2}L^{\dagger}L\right)  U\left(
t\right)  dt+\sum_{i=1}^{2m}L_{i}U\left(  t\right)  dA_{i}^{\dagger}\left(
t\right)  -\sum_{i=1}^{2m}L_{i}^{\dagger}U\left(  t\right)  dA_{i}\left(
t\right) \label{EQA501}\\
U\left(  0\right)   & =I\nonumber
\end{align}
where $H\left(  u\right)  $ is the controlled Hamiltonian. Adapted process
$A_{i}\left(  t\right)  $, $A_{i}^{\dagger}\left(  t\right)  $ and $Idt$
satisfy the It\^{o}'s multiplication rule. The creation process $A_{i}%
^{\dagger}\left(  t\right)  $ and the annihilation process $A_{i}\left(
t\right)  $ are diffusive.

The time evolution of the operator $X\in\mathcal{A}$ is defined as the
Markovian flow $j_{t}\left(  X\right)  =U^{\dagger}\left(  t\right)  \left(
X\otimes I\right)  U\left(  t\right)  $. The application of the It\^{o}'s
relations to $X\left(  t\right)  =j_{t}\left(  X\right)  $ allows us to obtain
the well-known Heisenberg-Langevin equations:%
\begin{equation}
dX_{k}\left(  t\right)  =\mathcal{L}_{t}\left[  X_{k}\left(  t\right)
\right]  dt+\sum_{i=1}^{d}\left[  X_{k}\left(  t\right)  ,L_{i}\right]
dA_{i}^{\dagger}\left(  t\right)  -\sum_{i=1}^{d}\left[  X_{k}\left(
t\right)  ,L_{i}^{\dagger}\right]  dA_{i}\left(  t\right) \label{EQA502}%
\end{equation}
where $\mathcal{L}_{t}\left[  X\right]  =j_{t}\left(  \mathcal{L}\left[
X\right]  \right)  $ stands for the time evolution of the
Gorini-Kossakovski-Sudarshan generator%
\begin{equation}
\mathcal{L}\left[  X\right]  =\frac{1}{2}\sum_{i=1}^{d}L_{i}^{\dagger}\left[
X,L_{i}\right]  -\frac{1}{2}\left[  X,L_{i}^{\dagger}\right]  L_{i}%
+\frac{\mathbf{i}}{\hbar}\left[  H_{sys}+H(u),X\right] \label{EQA503}%
\end{equation}
and is given by%
\begin{equation}
\mathcal{L}_{t}\left[  X\right]  =j_{t}\left(  \mathcal{L}\left[  X\right]
\right)  =\frac{1}{2}\sum_{i=1}^{d}L_{i}^{\dagger}\left(  t\right)  \left[
X\left(  t\right)  ,L_{i}\left(  t\right)  \right]  -\frac{1}{2}\sum_{i=1}%
^{d}\left[  X\left(  t\right)  ,L_{i}^{\dagger}\left(  t\right)  \right]
L_{i}\left(  t\right)  +\frac{\mathbf{i}}{h}\left[  H_{sys}+H\left(  u\right)
,X\left(  t\right)  \right] \label{EQA401}%
\end{equation}
In deriving $\left(  \ref{EQA401}\right)  $ note that $j_{t}\left(  \left[
X,Y\right]  \right)  =\left[  X\left(  t\right)  ,Y\left(  t\right)  \right]
$ and that $j_{t}\left(  XY\right)  =X\left(  t\right)  Y\left(  t\right)  $,
for arbitrary operators $X,Y\in\mathcal{A}$.

\subsection{Evolution of the Operator $\mathcal{L}_{t}\left[  X_{k}\right]  $}

\subsubsection{System Hamiltonian}

For the system Hamiltonian $H_{sys}=\mathbb{X}^{\intercal}\left(  R\otimes
I\right)  \mathbb{X}$ we first derive $\left[  H_{sys},X_{k}\right]
=H_{sys}X_{k}-X_{k}H_{sys}$. The identities $\left[  X_{k},X_{i}\right]
=\hbar S_{k,i}$ and $X_{j}X_{k}=\left[  X_{j},X_{k}\right]  +X_{k}X_{j}$
allows us to write the following relations:%
\begin{align*}
H_{sys}X_{k}  & =\frac{1}{2}\sum_{i,j}r_{i,j}X_{i}X_{j}X_{k}=\frac{1}{2}%
\sum_{i,j=1}^{2m}r_{i,j}\left(  \hbar S_{k,i}S_{j,k}X_{i}+X_{i}X_{k}%
X_{j}\right) \\
X_{k}H_{sys}  & =\frac{1}{2}\sum_{i,j}r_{i,j}X_{k}X_{i}X_{j}=\frac{1}{2}%
\sum_{i,j=1}^{2m}r_{i,j}\left(  \hbar S_{k,i}S_{k,i}X_{j}+X_{i}X_{k}%
X_{j}\right)
\end{align*}
and then%
\begin{equation}
\left[  H_{sys},X_{k}\right]  =\frac{1}{2}\sum_{i,j=1}^{2m}r_{i,j}\hbar
S_{k,i}\left(  S_{k,i}S_{j,k}X_{i}-S_{k,i}S_{k,i}X_{j}\right) \label{EQA403}%
\end{equation}
On the other hand%
\begin{equation}
\sum_{i,j=1}^{2m}r_{i,j}S_{k,i}X_{j}=\hbar\left(
\begin{array}
[c]{ccccc}%
S_{k,1}I & S_{k,2}I & S_{k,3}I & \cdots & S_{k,2m}I
\end{array}
\right)  \left(  R\otimes I\right)  \mathbb{X}\label{EQA404}%
\end{equation}
Similarly,%
\begin{gather}
\mathbf{i}\hbar\sum_{i,j=1}^{2m}r_{i,j}S_{j,k}X_{i}=\mathbb{X}^{\intercal
}\left(  R\otimes I\right)  \left(
\begin{array}
[c]{c}%
S_{1,k}I\\
S_{2,k}I\\
\vdots\\
S_{2m,k}I
\end{array}
\right)  =\label{EQA405}\\
=-\hbar\left(
\begin{array}
[c]{ccccc}%
S_{k,1}I & S_{k,2}I & S_{k,3}I & \cdots & S_{k,m}I
\end{array}
\right)  \left(  R^{T}\otimes I\right)  \mathbb{X}\nonumber
\end{gather}
For the sake of simplicity we use the notation $\left[  H_{sys},\mathbb{X}%
\right]  $ to refer to the stacking of operators $\left[  H_{sys}%
,X_{k}\right]  $. According to $\left(  \ref{EQA404}\right)  $ and $\left(
\ref{EQA405}\right)  $ it is straightforward that
\begin{equation}
\left[  H_{sys},\mathbb{X}\right]  =-\frac{1}{2}\hbar\left(  \mathbb{S}\left(
R+R^{T}\right)  \otimes I\right)  \mathbb{X}\label{EQA513}%
\end{equation}

\subsubsection{Controlled Hamiltonian}

We define the controlled Hamiltonian as%
\begin{equation}
H\left(  u\right)  =\frac{1}{2}\left(  u^{T}\left(  K^{T}+K^{\dagger}\right)
\otimes I\right)  \mathbb{X}\label{EQA504}%
\end{equation}
or more explicitly%
\begin{equation}
H\left(  u\right)  =\frac{1}{2}\sum_{i=1}^{2m}\left(  u^{T}\left(
K^{T}+K^{\dagger}\right)  \mathbf{e}_{i}\right)  X_{i}\label{EQA505}%
\end{equation}
For each $X_{k}\in\mathcal{A}$ in $\mathbb{X}$ we compute the bracket $\left[
H\left(  u\right)  ,X_{k}\right]  $. Since $H\left(  u\right)  $ is a linear
combination of operators $X_{i}$, and recalling that $\left[  X_{i}%
,X_{k}\right]  =\hbar S_{i,k}$, we can write%
\[
\left[  H\left(  u\right)  ,X_{k}\right]  -\frac{1}{2}\sum_{i=1}^{2m}\hbar
S_{k,i}e_{i}^{T}\left(  K+K^{\ast}\right)  uI
\]
and in compact form%
\[
\left[  H\left(  u\right)  ,\mathbb{X}\right]  =-\frac{\hbar}{2}\left(
\mathbb{S}\left(  K+K^{\ast}\right)  \otimes I\right)  \left(  u\otimes
I\right)
\]
Defining $B=\frac{1}{2}\left(  \mathbb{S}\left(  K+K^{\ast}\right)  \otimes
I\right)  $ and $v\left(  t\right)  =\left(  u\left(  t\right)  \otimes
I\right)  $, we get the control term $Bv\left(  t\right)  $ in the SDE.

\subsubsection{Coupling terms}

The coupling operator $L_{i}$ for the i-th measurement channel depends on the
state $\mathbb{X}$:%
\[
L_{i}=\sum_{i=1}^{2m}\Gamma_{il}X_{l}%
\]
We begin writing the brackets%
\begin{equation}
\left[  X_{k},L_{i}\right]  =\sum_{j=1}^{2m}\Gamma_{i,j}\left[  X_{k}%
,X_{j}\right]  =\hbar\sum_{j=1}^{2m}\Gamma_{i,j}S_{k,j}I\label{EQA507}%
\end{equation}

\begin{equation}
\left[  L_{i}^{\dagger},X_{k}\right]  =\sum_{j=1}^{2m}\Gamma_{i,j}^{\ast
}\left[  X_{j}^{\dagger},X_{k}\right] \label{EQA506}%
\end{equation}
The bracket $\left[  X_{j}^{\dagger},X_{k}\right]  $ in $\left(
\ref{EQA506}\right)  $ is a little more involved and deserves more attention.
First of all,%
\begin{equation}
\left[  X_{j}^{\dagger},X_{k}\right]  =\left\{
\begin{array}
[c]{cc}%
\left[  X_{j+m},X_{k}\right]  =\hbar S_{j+m,k} & \text{, if }j\leq m\\
\left[  X_{j}^{\dagger},X_{k}\right]  =\left[  \left(  X_{j-m}^{\dagger
}\right)  ^{\dagger},X_{k}\right]  =\hbar S_{j-m,k} & \text{, if }j>m
\end{array}
\right. \label{EQA508}%
\end{equation}
Secondly, from $\left(  \ref{EQA508}\right)  $ we observe that the effect of
$\left[  X_{j}^{\dagger},X_{k}\right]  $ is to exchange the row blocks in the
partitioned matrix $\mathbb{S}$, i.e. $\left[  X_{j}^{\dagger},X_{k}\right]
=\hbar\left(  \mathbb{JS}\right)  _{j,k}I$. By exploiting the identities
$\left(  \mathbb{JS}\right)  ^{T}=\mathbb{S}^{T}\mathbb{J}^{T}=-\mathbb{SJ}$,
the bracket is conveniently written as $\left[  X_{j}^{\dagger},X_{k}\right]
=-\hbar\left(  \mathbb{SJ}\right)  _{j,k}I$.

Introducing this information into $\left(  \ref{EQA506}\right)  $ and $\left(
\ref{EQA507}\right)  $ yields:%

\begin{equation}
\sum_{i=1}^{d}L_{i}^{\dagger}\left[  X_{k},L_{i}\right]  =\hbar\sum_{i=1}%
^{d}\sum_{j,l=1}^{2m}\Gamma_{i,j}\Gamma_{i,l}^{\ast}S_{k,j}X_{l}^{\dagger
}\label{EQA509}%
\end{equation}%
\begin{equation}
\sum_{i=1}^{d}\left[  L_{i}^{\dagger},X_{k}\right]  L_{i}=-\hbar\sum_{i=1}%
^{d}\sum_{j,l=1}^{2m}\Gamma_{i,j}^{\ast}\Gamma_{i,l}\left(  \mathbb{SJ}%
\right)  _{j,k}X_{l}\label{EQA510}%
\end{equation}
Finally,%
\begin{equation}
\sum_{i=1}^{d}L_{i}^{\dagger}\left[  X_{k},L_{i}\right]  =\hbar\sum_{i=1}%
^{d}\sum_{j,l=1}^{2m}S_{k,j}\Gamma_{j,i}^{T}\Gamma_{i,l}^{\ast}X_{l}^{\dagger
}=\hbar\sum_{l=1}^{2m}\left(  \mathbb{S}\Gamma^{T}\Gamma^{\ast}\right)
_{kl}X_{l}^{\dagger}\label{EQA511}%
\end{equation}%
\begin{equation}
\sum_{i=1}^{d}\left[  L_{i}^{\dagger},X_{k}\right]  L_{i}=-\hbar\sum
_{l=1}^{2m}\left(  \mathbb{SJ}\Gamma^{\dagger}\Gamma\right)  _{kl}%
X_{l}\label{EQA512}%
\end{equation}
and in compact form%
\begin{equation}
\frac{1}{2}\left(  \sum_{i=1}^{d}L_{i}^{\dagger}\left[  X,L_{i}\right]
-\left[  X,L_{i}^{\dagger}\right]  L_{i}\right)  =\frac{\hbar}{2}%
\mathbb{S}\left(  \left(  \Gamma^{T}\Gamma^{\ast}\mathbb{J}+\mathbb{J}%
\Gamma^{\dagger}\Gamma\right)  \otimes I\right)  \mathbb{X}\label{EQA514}%
\end{equation}
From $\left(  \ref{EQA513}\right)  $ and $\left(  \ref{EQA514}\right)  $ we
define
\[
A=\frac{\hbar}{2}\left(  \mathbb{S}\left(  R+R^{T}+\left(  \Gamma^{T}%
\Gamma^{\ast}\mathbb{J}+\mathbb{J}\Gamma^{\dagger}\Gamma\right)  \right)
\otimes I\right)
\]
which corresponds to the term $A\mathbb{X}$ in the SDE.

\subsubsection{Terms of Uncertainty}

The uncertainty terms in $\left(  \ref{EQA502}\right)  $ can be computed by
resorting to $\left(  \ref{EQA506}\right)  $, $\left(  \ref{EQA507}\right)  $
and $\left(  \ref{EQA508}\right)  $:%
\[
\sum_{i=1}^{d}\left[  X_{k}\left(  t\right)  ,L_{i}\right]  dA_{i}^{\dagger
}\left(  t\right)  =\sum_{i=1}^{d}\sum_{j=1}^{2m}\Gamma_{i,j}\left[
X_{k},X_{j}\right]  dA_{i}^{\dagger}\left(  t\right)  =\hbar\sum_{i=1}^{d}%
\sum_{j=1}^{2m}S_{k,j}\left(  \Gamma^{T}\right)  _{j,i}dA_{i}^{\dagger}\left(
t\right)
\]

\begin{equation}
\sum_{i=1}^{d}\left[  L_{i}^{\dagger},X_{k}\right]  dA_{i}\left(  t\right)
=\hbar\sum_{i=1}^{d}\sum_{j=1}^{2m}\left(  \mathbb{JS}\right)  _{k,j}%
\Gamma_{j,i}^{\dagger}dA_{i}\left(  t\right)
\end{equation}
Calling $\mathbb{A}$ and $\mathbb{A}^{\dagger}$to the stacking of operators
$A_{i}$ and $A_{i}^{\dagger}$, the noise increment in the state equation is
given by%
\[
dW\left(  t\right)  =\hbar\left(  \mathbb{S}\Gamma^{T}\mathbb{J}\otimes
I\right)  d\mathbb{A}^{\dagger}\left(  t\right)  +\hbar\left(  \mathbb{JS}%
\Gamma\otimes I\right)  d\mathbb{A}\left(  t\right)
\]

\subsubsection{Output Equation}

The field quadrature for the i-th output channel is given by $A_{i}%
+A_{i}^{\dagger}$. The output in the channel $i$ is considered as the weak
measurement $Y_{i}\left(  t\right)  =U^{\dagger}\left(  t\right)  \left(
\left(  A_{i}+A_{i}^{\dagger}\right)  \otimes I\right)  U\left(  t\right)  $.
The application of the It\^{o} formula to $Y_{i}\left(  t\right)  $ yields:%
\[
dY_{i}\left(  t\right)  =\left(  L_{i}+L_{i}^{\dagger}\right)  dt+dA_{i}%
+dA_{i}^{\dagger}%
\]
which can be compactly written as%
\[
dY\left(  t\right)  =\left(  \left(  \Gamma+\Gamma^{\ast}\right)  \otimes
I\right)  \mathbb{X}dt+\left(  d\mathbb{A}\left(  t\right)  +d\mathbb{A}%
^{\dagger}\left(  t\right)  \right)
\]
Note that%
\begin{align*}
L_{i}+L_{i}^{\dagger}  & =\sum_{l=1}^{2m}\Gamma_{il}X_{l}+\Gamma_{il}^{\ast
}X_{l}^{\dagger}=\\
& =\sum_{l=1}^{m}\Gamma_{il}X_{l}+\sum_{l=m+1}^{2m}\Gamma_{il}^{\ast}X_{l}+\\
& +\sum_{l=m+1}^{2m}\Gamma_{il}X_{l}+\sum_{l=1}^{m}\Gamma_{il}^{\ast}X_{l}%
\end{align*}

\section*{Acknowledgments}
JMV is supported by Ministerio de Economia y Competitividad FIS2015-69512-R
and Programa de Excelencia de la Fundacion Seneca 19882/GERM/15. 

\section*{Data Availability}
The data that support the findings of this study are available from the corresponding author upon reasonable request.

\bibliographystyle{siamplain}

\end{document}